\documentclass[authoryear,review,12pt,pdf]{elsarticle}
\usepackage[toc,page]{appendix}
\usepackage{adjustbox}
\usepackage{amssymb}
\usepackage{amsmath}
\usepackage{array}
\usepackage{blindtext}
\usepackage{booktabs} 

\usepackage[labelfont=bf]{caption}
\DeclareCaptionLabelFormat{mycaption}{#1 #2.}
\usepackage{color}
\usepackage[dvipsnames]{xcolor}
\usepackage{csquotes}
\usepackage{float}
\usepackage{geometry} 
\usepackage{graphicx}
\usepackage[colorlinks=true,linkcolor=black, citecolor=blue, urlcolor=blue]{hyperref}
\usepackage{lipsum}
\usepackage{cite}
\usepackage{gensymb}
\usepackage{lscape}
\usepackage{lineno}
\usepackage{mathtools}
\usepackage{makecell}
\usepackage{multirow}
\usepackage{pifont}
\usepackage{pdflscape}
\usepackage{rotating}
\usepackage{soul}
\setstcolor{red}
\usepackage{tabularx,booktabs}
\usepackage{threeparttable}
\usepackage{textcomp}
\usepackage{tikz} \usetikzlibrary{shapes,arrows}
\usepackage{url}
\usepackage{indentfirst}
\usepackage{varioref}
\usepackage{wrapfig}
\usepackage[dvipsnames]{xcolor}
\usepackage{wrapfig}
 \usepackage{subfigure}
 \usepackage[toc,page]{appendix}
\definecolor{dblue}{rgb}{0.16,0.49,0.66}
\usepackage{siunitx}
\newcolumntype{P}[1]{>{\centering\arraybackslash}p{#1}}
\newcolumntype{M}[1]{>{\centering\arraybackslash}m{#1}}

\DeclareMathAlphabet {\mathbfit}{OML}{cmm}{b}{it}

\usepackage{booktabs,array,dcolumn}
\DeclareUnicodeCharacter{2212}{-}
\DeclareUnicodeCharacter{0301}{'}

\usepackage{graphicx} 
\usepackage{multirow} 

\begin{document}
\hypersetup{allcolors = [rgb]{0.16,0.49,0.66}}

\begin{frontmatter}

\title{Global Estimates of L-band Vegetation Optical Depth and Soil Permittivity over Snow-covered Boreal Forests and Permafrost using SMAP Satellite Data}
\author[add1]{Divya Kumawat} 
\author[add1]{Ardeshir Ebtehaj,\corref{cor1}}\ead{ebtehaj@umn.edu}
\author[add2]{Mike Schwank}
\author[add3]{Jean-Pierre Wigneron}
\author[add3]{ Xiaojun Li}

\cortext[cor1]{Corresponding author}
\address[add1]{Saint Anthony Falls Laboratory, Department of Civil Environmental and Geo-Engineering, University of Minnesota, Minneapolis, MN 55414, United States}
\address[add2]{Federal Research Institute for Forrest, Snow and Landscape Research WSL, Zürcherstrasse 111, Birmensdorf 8903, Switzerland}
\address[add3]{Institut national de la recherche agronomique, UMR1391 ISPA, F-33140, Villenave d'Ornon, France}
\vspace{-2mm}
\begin{abstract} \vspace{-2mm}
This article expands the tau-omega model to properly simulate L-band microwave emission of the soil-snow-vegetation continuum through a closed-form solution of Maxwell's equations, considering the intervening dry snow layer as a loss-less medium. The feasibility and uncertainty of retrieving vegetation optical depth (VOD) and ground permittivity, given the noisy L-band brightness temperatures with 1~\si{K} (1-sigma), are demonstrated through controlled numerical experiments. For moderately dense vegetation canopy and a range of 100--400 \si{kg.m^{-3}} snow density, the standard deviation of the retrieval errors is 0.1 and 3.5 for VOD and ground permittivity respectively. Using L-band observations from the Soil Moisture Active Passive (SMAP) satellite, a new data set of global estimates of VOD and ground permittivity are presented over the Arctic boreal forests and permafrost areas during winter months. In the absence of dense ground-based observations of ground permittivity and VOD, the retrievals are causally validated using dependent variables including above-ground biomass, tree height, and net ecosystem exchange. Time-series analyses promise that the new data set can expand our understanding of the land-atmosphere interactions and exchange of carbon fluxes over the Arctic landscape. 

\end{abstract}


\end{frontmatter}

\section{Introduction}
\label{sec:I}

Space-time monitoring of vegetation water content (VWC) on a global scale is of utmost importance in understanding the impacts of climate change on vegetation biomes, phenology, and ecosystem interactions \citep{richardson2013climate}. Global forests are in a dynamic state of change, with 2.3 million square kilometers lost and 1.3 million square kilometers gained from 2000 to 2020 \citep{potapov2022global}. The changes in the patterns of vegetation phenology have a significant impact on the exchange rates of radiative energy, water, and greenhouse gases between the land and atmosphere, influencing the global and regional climate systems and carbon cycle \citep{piao2020characteristics}. Global warming and the Arctic sea ice decline have resulted in changes in precipitation \citep{Tamang2019} and early onset of the growing season over the northern hemisphere (NH) that increased peak annual greenness in Arctic tundra biomes by up to 0.79\%/yr \citep{jia2009vegetation}, in terms of the normalized difference vegetation index (NDVI). Furthermore, the transition zone between the boreal forest and the Arctic tundra is experiencing a shift, with noticeable poleward latitudinal advance rates ranging from around 10~\si{m.year^{-1}} in Canada to as high as 100~\si{m.year^{-1}} in Western Eurasia \citep{rees2020subarctic}.

Snow-covered boreal forests contain one-third of the world's terrestrial carbon pool and thus play a critical role in the global carbon cycle \citep{bradshaw2015global}. This pool in the Arctic permafrost ($\sim$1,460-1,600 billion tons, \citet{hugelius2014estimated}) can be gradually released to the atmosphere in response to potential future thawing processes \citep{jin2021impacts} and accelerated soil microbial activities \citep{romanovsky2010thermal}. The dynamics of boreal forests and permafrost regions are tightly connected. Predictive modeling suggests that slow and steady thawing of permafrost will release around 200 billion tonnes of carbon over the next 300 years under the current warming trends \citep{mcguire2018dependence}. This additional flux can lead to increased greening that might offset its impacts on the ecosystem \citep{wei2021plant}; however, this competing and highly complex feedback is yet not well understood \citep{schuur2022permafrost}. To improve our understanding of these changes, it seems imperative to use satellite observations of the soil-snow-vegetation continuum over the Arctic tundra, boreal forests, and permafrost. 

Spaceborne observations at visible and infrared wavelengths have facilitated the monitoring of vegetation dynamics and their implications on global patterns of surface carbon fluxes \citep{jung2011global} since the early 1980s -- through the NDVI and the enhanced vegetation index (EVI). While these indices can effectively measure photosynthetic activity and leaf area index (LAI), they are not always a reliable indicator of total above-ground biomass (AGB) except in areas with low vegetation density \citep{todd1998biomass}. Moreover, these optical indices become highly uncertain over high latitudes due to the presence of snow cover and low light illumination, especially during the winter.

Unlike NDVI and EVI, the vegetation optical depth (VOD, $\tau$) obtained from passive microwave observations can provide important complementary information on the state and temporal changes of VWC. Studies found a strong correlation between L-band VOD and AGB on a global scale \citep{liu2015recent,rodriguez2018evaluation,brandt2018satellite,vittucci2019vegetation,frappart2020global,wigneron2021alternate}. These studies showed that L-band VOD can be successfully used as a surrogate variable to estimate tropical forest biomass at a continental scale, making it an important variable to model the global carbon stock and cycle \citep{houghton2005aboveground}.

The European Space Agency's (ESA) Soil Moisture and Ocean Salinity (SMOS) satellite \citep{kerr2010smos} and the National Aeronautics and Space Administration's (NASA) Soil Moisture Active Passive (SMAP) satellite \citep{entekhabi2010soil} have played a critical role in providing global estimates of VOD and soil moisture through L-band radiometry. However, currently, the retrievals of these variables are limited or not available where snow covers a significant portion of the radiometric field of view. The reason is that current emission models, used in satellite global data products, do not account for the effects of snow cover. It is important to note that, the soil is not always frozen below snowpack and can remain partially unfrozen for several weeks even after the temperature of the soil goes below the freezing point depending on the soil mineralogy \citep{sutinen2008effect}. Using numerous reanalysis data sets, it was also shown that more than 30\% of the time, the soil below NH's snowpack can be unfrozen \citep{gao2022variability}, especially during the accumulation season.   

\begin{figure}[t]

\centering\includegraphics[width=1\linewidth]{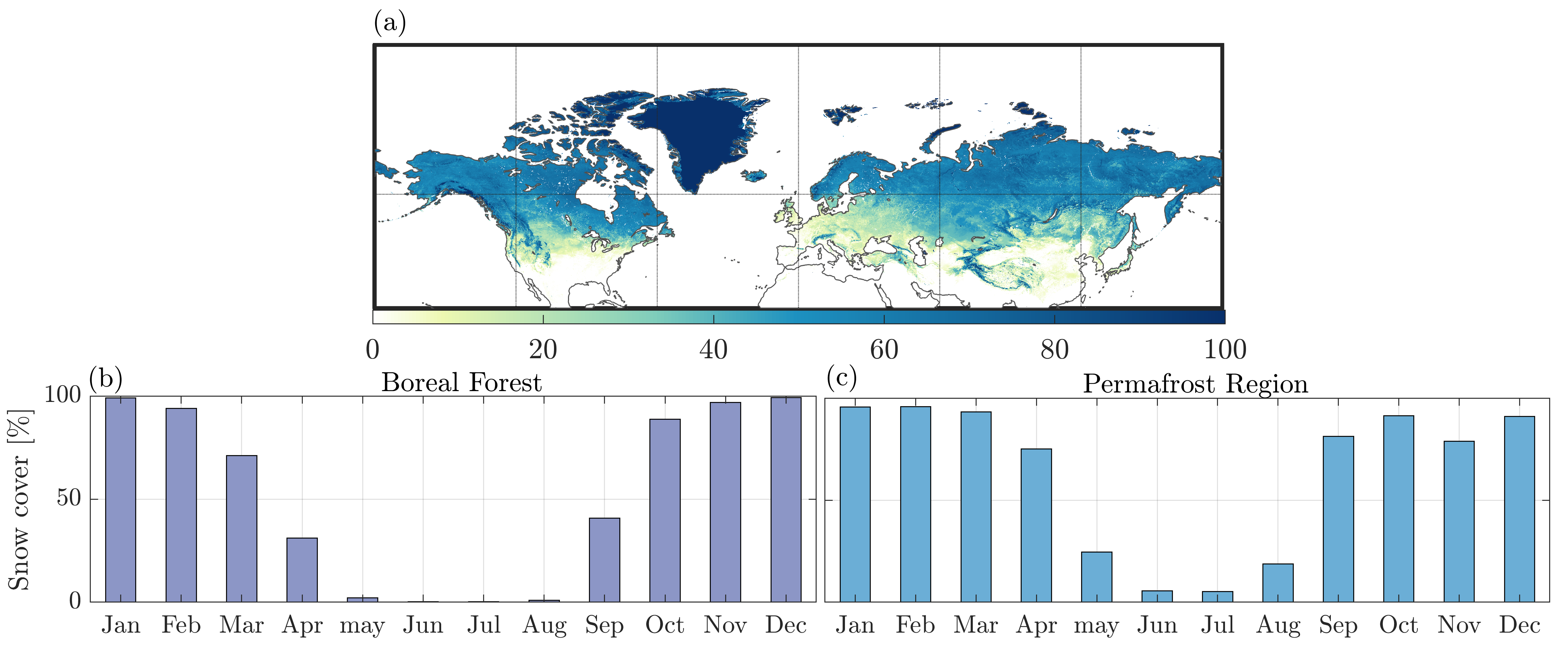}
\caption{(a) Mean annual percentage of snow cover in 2017 using level-III global monthly gridded products (MYD10CM) at 0.05$^\circ$ resolution from the Moderate Resolution Imaging Spectroradiometer (MODIS) instrument onboard the NASA’s Aqua satellite, as well as (b) monthly snow-cover percentage over the boreal forest and (c) permafrost regions. The boundary of the boreal forests and permafrost is from \citet{potapov2008combining} and \citet{obu2018gtmn}, respectively. The permafrost is delineated using the exceedance probability of 50\% capturing the presence of continuous and discontinuous permafrost at a 1~km grid size.} \label{fig:1}
\end{figure}

Fig.~\ref{fig:1}\,a shows the annual spatial NH's snow-cover fraction in 2017. The monthly values, averaged over the boreal forests and permafrost regions, are shown in Fig.~\ref{fig:1}\,b,c, respectively.  It is evident that both boreal forests (permafrost) are covered by snow more than 50\% of the time from October (September) to March (April). This observation indicates that the availability of satellite-based VOD ($\tau$) and ground permittivity ($\varepsilon_{\rm g}$) data is severely limited for over six months of the year over these important land surface types.

The lack of L-band satellite data on ground permittivity and VOD over snow-covered surfaces is primarily due to the complex effects of snow on soil and vegetation emissions. Although dry snow is a low-loss medium at L-band, its dielectric constant varies significantly as a function of its physical characteristics mainly density and liquid water content \citep{matzler1984microwave,schwank2015snow}. This dependency will affect the refraction of upwelling (downwelling) soil (vegetation) at the interfaces of soil, snow, and vegetation. Experimental and theoretical studies \citep{lemmetyinen2016snow,naderpour2017snow,kumawat2022passive} have shown that the presence of dry snow may increase surface emissions, and a failure to account for its signal can lead to an underestimation of ground permittivity by 30--40\%, with a higher margin of error for a denser snowpack. These effects are not currently accounted for in operational algorithms and available data sets from the SMAP and SMOS satellites. 

To address the existing limitations, two emission models have been developed including the zeroth-order tau-omega-snow \citep[TO-snow,][]{kumawat2022passive}, and the first-order two-stream \citep[2-S,][]{schwank2014model} model. The TO-snow model expands on the classic tau-omega model \citep[TO,][]{mo1982model}, used for official SMAP retrievals. The first version of the TO-snow employs the dense media radiative transfer theory \citep[DMRT,][]{tsang2000dense} for computation of the snow dielectric constant. The surface reflection of the downwelling vegetation emission was accounted for through a coherent two-layer composite model that represents multiple reflections within the snow layer while vegetation is considered to be a weakly scattering medium. In contrast, the 2-S model, based on the microwave emission model of a layered snowpack \citep[MEMLS,][]{matzler1998improved,wiesmann1999microwave}, takes into account multiple scattering and reflection for the vegetation layer. 

Building upon the previous work, a modified version of the zeroth-order TO-snow model is proposed in this paper and applied to the SMAP data on a global scale to investigate the following main research questions. (1) What are the impacts of snow density, soil roughness, and ground permittivity on the observed surface emissions for different VOD values? (2) what are the expected uncertainties in the VOD and ground permittivity retrievals over snow-covered areas and how do they depend on soil roughness and snow density? (3) What are the correlations between the retrieved ground permittivity and VOD with various vegetation proxies and net ecosystem exchange (NEE) over the NH's boreal forests and permafrost? 

The paper is organized as follows. Section~\ref{sec:II} explains the structure of the used emission model followed by the inversion method. The used data sets are explained in Section~\ref{sec:III}. Implementation and results are presented in Section~\ref{sec:IV}. Section~\ref{sec:V} concludes and discusses current shortcomings and potential extensions of the research. 

\section{Methodology}
\label{sec:II}

\subsection{TO-snow emission model}
\label{sec:II.1}

In the TO-snow emission model, the simulated brightness temperatures $T_{\rm B}^p$ at polarization $p$ comprise three components: (i) upwelling soil emission, (ii) reflected downwelling vegetation emission with multiple reflections and refractions at the soil-snow and snow-vegetation interfaces undergoing single-way (zeroth-order) attenuation by the vegetation canopy, and (iii) the upwelling emission by the canopy layer. In the first version, the upwelling soil emission has been computed using the DMRT model \citep{tsang2000dense}. However, in the modified version, to reduce the complexity and computation time, the soil emission through the dry snow is simulated using a coherent two-layer composite model, obtained through a closed-form solution of the Maxwell equations, that can account for multiple reflections within the snow layer \citep{ulaby2014microwave} as a lossless intervening medium. Similar to the TO model, the overlying canopy is considered to be a weakly scattering medium and the extinction of surface emission is represented through the one-way vegetation transmissivity \citep{mo1982model}. 

The model simulates the brightness temperature $T_{\rm B}^p$ at the top of the canopy as follows:

\begin{equation}
T_{\rm B}^p ={\overbrace{T_{\rm g} e^{\,p} \gamma}^{(1)}+\overbrace{T_{\rm c}(1-\omega)(1-\gamma)r^{p}\gamma}^{(2)}+ \overbrace{T_{\rm c}(1-\omega)(1-\gamma)}^{(3)}}. \label{eq:01}
\end{equation}
where the upwelling surface emission is $T_{\rm g}e^{p}\gamma$, in which $T_{\rm g}$ is the effective ground temperature, $e^p$ denotes the incoherent component of effective emissivity of the soil-snow system, and $\gamma = \exp(-\tau \sec \alpha_i)$ represents the vegetation transmissivity as a function of VOD ($\tau$) at observation angle $\alpha_i$ relative to nadir. The upwelling vegetation emission is $T_{\rm c}(1-\omega)(1-\gamma)$ where $T_{\rm c}$ is the canopy temperature and $\omega$ is the vegetation single scattering albedo. The downwelling vegetation emission reflected by the soil-snow layers is $T_{\rm c}(1-\omega)(1-\gamma)r^p\gamma$, in which $r^p$ is the incoherent component of effective surface reflectivity of the soil-snow system.

To compute the effective emissivity of the soil-snow system, we used a coherent two-layer composite reflection model \citep[][p.~64]{ulaby2014microwave}, considering a single-layer representation of dry snowpack, with thickness $d_{\rm s}$,  without accounting for the effects of snow layering. In this formalism, we assume that the effective index of refraction of the vegetation layer is close to 1 and thus the interface of vegetation and air is a diffused boundary without any distinct refraction, commonly known as a soft layer. Consequently, using the propagation-matrix method \citep{tsang1985theory}, the vertically (V-pol) and horizontally (H-pol) polarized coherent (effective) surface emissivity $e^p_{\rm coh}$ and reflectivity $r^p_{\rm coh}$ at the snowpack surface can be obtained as follows:
\begin{equation}
r^p_{\rm coh} = \left\lvert\frac{\xi_{\rm cs}^p+\tilde{\xi}_{\rm sg}^p\,e^{-2\gamma_{\rm s}d_{\rm s}\cos\alpha_{\rm s}}}{1+\xi_{\rm cs}^p\,\tilde{\xi}_{\rm sg}^p\,e^{-2\gamma_{\rm s}d_{\rm s}\cos\alpha_{\rm s}}}\right\rvert^2 \quad \text{and} \quad
e^p_{\rm coh} = 1- \left\lvert\frac{\tilde{\xi}_{\rm sg}^p+\xi_{\rm cs}^p\,e^{-2\gamma_{\rm s}d_{\rm s}\cos\alpha_{\rm s}}}{1+\xi_{\rm cs}^p\,\tilde{\xi}_{\rm sg}^p\,e^{-2\gamma_{\rm s}d_s\cos\alpha_{\rm s}}}\right\rvert^2, \label{eq:02} 
\end{equation}

\noindent where $\xi^p_{\rm cs}$ is the Fresnel reflection coefficient of the canopy-snow (cs) interface, $\tilde{\xi}_{\rm sg}^p$ denotes the rough reflection coefficient at the snow-ground (sg) interface, $\gamma_{\rm s}$ and $\alpha_{\rm s}$ are the complex propagation constant and angle within the snow layer, respectively.

The rough surface soil reflection coefficient $\tilde{\xi}_{\rm sg}^p$ is related to its smooth counterpart $\xi_{\rm sg}^p$ via $|\tilde{\xi}_{\rm sg}^p|^2=|{\xi_{\rm sg}^p}|^2\exp(-h\cos^2\alpha_{\rm s})$, where $h$ is the surface soil roughness parameter, assumed to be linearly related to the root-mean-squared surface height in the well-known Q-H roughness model \citep{choudhury1979effect, wang1983passive}. The field reflection coefficients $\xi_{\rm cs}^p$ and ${\xi_{\rm sg}^p}$ are polarization dependent and are calculated using the intrinsic impedance of the soft air-canopy layer ($\eta_{\rm c}$), snow ($\eta_{\rm s}$), and ground ($\eta_{\rm g}$) -- through the Fresnel equations as follows:

\begin{equation}
\begin{aligned}
    \xi_{\rm cs}^{\rm H} & = \frac{\eta_{\rm s} \cos\alpha_{\rm i} - \eta_{\rm c}\cos\alpha_{\rm s}}{\eta_{\rm s} \cos\alpha_{\rm i} + \eta_{\rm c}\cos\alpha_{\rm s}} \\
    \xi_{\rm sg}^{\rm H} & = \frac{\eta_{\rm g} \cos\alpha_{\rm s} - \eta_{\rm s}\cos\alpha_{\rm g}}{\eta_{\rm g} \cos\alpha_{\rm s} + \eta_{\rm s}\cos\alpha_{\rm g}}, \\
    \xi_{\rm cs}^{\rm V} & = \frac{\eta_{\rm s} \cos\alpha_{\rm s} - \eta_{\rm c}\cos\alpha_{\rm i}}{\eta_{\rm s} \cos\alpha_{\rm s} + \eta_{\rm c}\cos\alpha_{\rm i}}, \\ 
    \xi_{\rm sg}^{\rm V} & = \frac{\eta_{\rm g} \cos\alpha_{\rm g} - \eta_{\rm s}\cos\alpha_{\rm s}}{\eta_{\rm g} \cos\alpha_{\rm g} + \eta_{\rm s}\cos\alpha_{\rm s}}, \label{eq:03}
\end{aligned}
\end{equation}
where $\alpha_{\rm g}$ is the observation angle at the snow-ground interface relative to the nadir, $\alpha_{\rm s}$ and $\alpha_{\rm g}$ are computed using Snell's law. Here we compute the relative permittivity of the loss-less dry snow (ds) as follows \citep{hallikainen1986dielectric,matzler2006thermal}:

\begin{equation}
     \varepsilon_{\rm ds}^{'} = 
\begin{cases}
    1+1.4667 v_{\rm i} + 1.435 v_{\rm i}^2 & 0\leq v_{\rm i}\leq 0.45\\
    (1+0.4759 v_{\rm i})^3               & v_{\rm i} > 0.45,
\end{cases} \\
\end{equation}
where $v_{\rm i} = \rho_{\rm s}/\rho_{\rm i}$ is the volume fraction of ice, $\rho_{\rm s}$ is the mean-mass density of the snowpack, and ice density is considered as $\rho_{\rm i}=0.9167$~\si{g.cm^{-3}}.

\begin{figure}[t!]
\centering\includegraphics[clip, trim=1.2cm 0.1cm 2.8cm 0.9cm,width=0.6\linewidth]{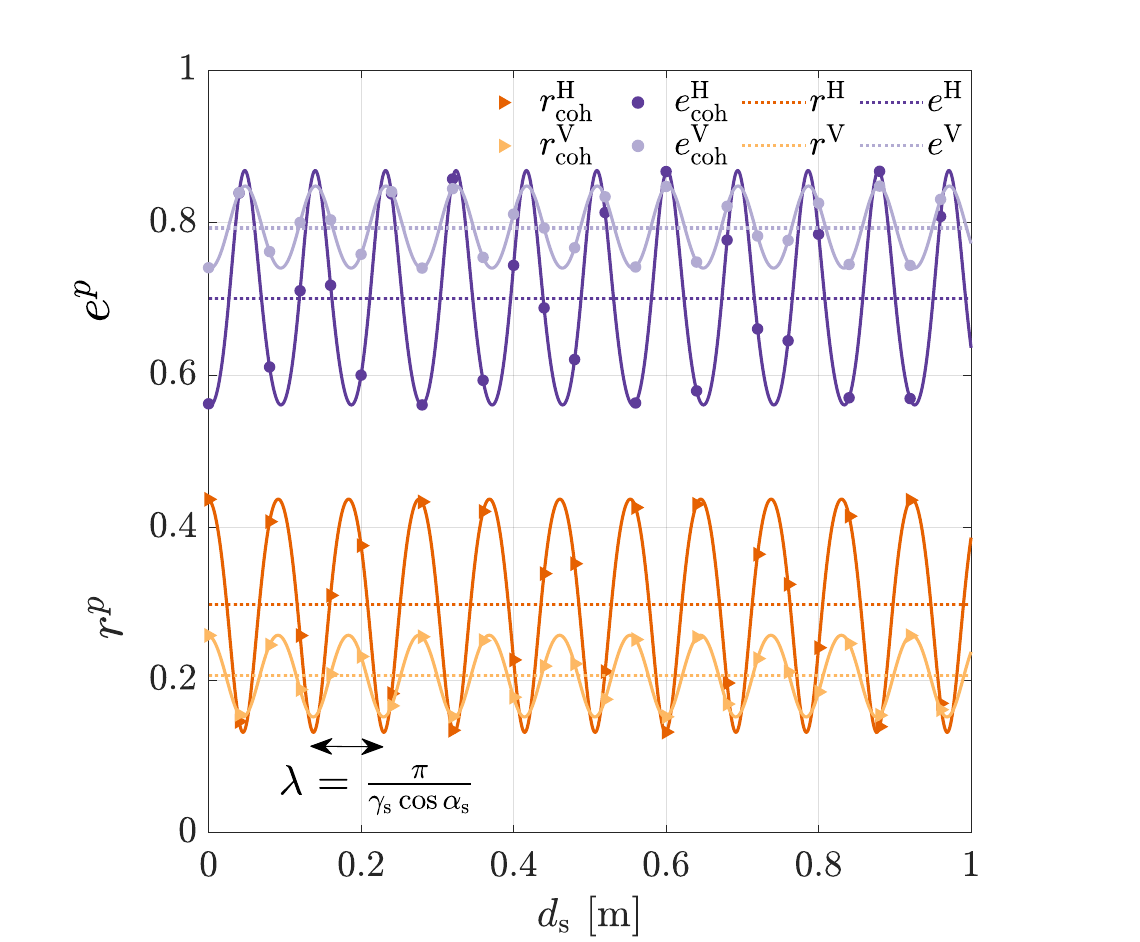}
\caption{Variation in the incoherent (mean, dashed lines) and coherent (oscillatory, solid lines) components of effective surface reflectivity $r^p$ and emissivity $e^p$ of soil-snow system with respect to the depth of snowpack $d_{\rm s}$ at density $\rho_{\rm s}$ = 400 \si{kg.m^{−3}}, ground permittivity $\varepsilon_{\rm g} = 20$, observation angle $\alpha_{\rm i} =  40^\circ$, as well as ground and canopy temperatures of $T_{\rm g}=275$~\si{K} and $T_{\rm c}=265$~\si{K}. The surface roughness parameter is set to $h=0.15$ and vegetation single scattering albedo is $\omega = 0.07$.} \label{fig:2}
\end{figure}

It is shown in Fig.~\ref{fig:2} that the response of $r_{\rm coh}^p$ and $e_{\rm coh}^p$ to $d_{\rm s}$ is oscillatory with no damping and a period of $ \pi/(\gamma_{\rm s}\cos\alpha_{\rm s})$ -- considering the dry snowpack as a lossless medium with a zero attenuation constant and a non-zero phase constant $\beta_{\rm s} \approx 2\pi/(\lambda_0\sqrt{\varepsilon_{\rm ds}^{'}})$, where $\lambda_0$ is the wavelength in the free-space. As is evident, if $d_{\rm s}$ varies more than 10~\si{cm}, within the field of view of the radiometer, it is reasonable to assume that the surface emission is incoherent with respect to depth \citep{naderpour2017snow} and can be represented as average values of the coherent counterpart as follows -- as derived in \ref{A1}:
\begin{equation}
 \begin{aligned}
     r^{\,p} & = 1+ \text{sgn}(|\xi_{\rm cs}^p| |\tilde{\xi}_{\rm sg}^p|-1)\left(\frac{|\xi_{\rm cs}^p|^2 + |\tilde{\xi}_{\rm sg}^p|^2-|\xi_{\rm cs}^p|^2 |\tilde{\xi}_{\rm sg}^p|^2-1}{|\xi_{\rm cs}^p|^2 |\tilde{\xi}_{\rm sg}^p|^2-1}\right) \\
    e^{\,p} & = 1-r^{\,p},
\end{aligned} \label{eq:05}
 \end{equation}
where $\text{sgn}(\cdot)$ denotes the signum function.

\subsection{2-S emission model}
\label{sec:II.2}

The 2-S emission \citep{schwank2014model} model can be used to simulate L-band microwave emission from snow-covered vegetated surfaces. This emission model is based on parts of MEMLS \citep{wiesmann1999microwave}, and for the application to the L-band it is assumed that absorption and volume scattering in dry snow can be neglected. Similar to the TO-snow model, snow is characterized by its permittivity, which is controlled by the dry snow density. Once the interface reflectivity values are known, the Kirchhoff coefficients associated with the canopy, snow, and ground layers are computed to derive the resultant brightness temperature. In order to compute the Kirchhoff coefficients, the emission model balances the up- and down-welling electromagnetic energy fluxes propagated into each layer by taking into account the conservation of energy at the layer interfaces. Unlike the TO-snow model, the 2-S emission model takes into account internal volume scattering in the vegetation layer derived from a six-flux approach however ignores the multiple reflections and refractions within the snow layer \citep{matzler2006thermal}. Furthermore, the contribution from sky radiation is included in the 2-S emission model.
 
\subsection{Inverse model}
\label{sec:II.3}

In this study, both horizontal and vertical polarized brightness temperatures are used \citep{ebtehaj2019physically,gao2020spatially} to estimate $\tau$ and $\varepsilon_{\rm g}$, based on least-squares minimization of the difference between the output of the emission model $f^p(\boldsymbol{\phi})$ and observations of the brightness temperature $\mathbf{y}_{T_{\rm B}}^p$, as follows:
\begin{equation}
\begin{split}
\phi^{*}= & \underset{\phi}{\text{argmin}\,}{\sum_p \left(\mathbf{y}_{T_{\rm B}}^p -f^p(\boldsymbol{\phi})\right)^2} + \mu(\tau-\tau_{0})^2\\ & \text{subject to} \quad \boldsymbol{\phi_l \leq \phi \leq \phi_u}, \label{eq:06}
\end{split}
\end{equation} 
where $\boldsymbol{\phi}$ =$\left(\tau,\,\varepsilon_{\rm g}\right)$, $f^p(\cdot)$ denotes a functional representation of the emission model at polarization $p$, $\mu>0$ is a regularization parameter, and $\tau_0$ is the VOD climatology, obtained from 15 years of MODIS NDVI data \citep[MOD13C1,][]{didan2015mod13c1}. The Tikhonov regularization term was applied to regress the retrievals slightly towards the VOD climatology as suggested in the latest version of the SMAP baseline algorithm \citep{chaubell2021regularized, chaparro2022robustness}. The upper $\phi_u$ and lower bounds $\phi_l$ of the parameters can be adjusted based on a priori knowledge from ground-based observations or reanalysis data.   

Using the regularization term in Eq.~\ref{eq:06} may not be the best approach as NDVI is not a proper representation of the VWC throughout the canopy layer. However, we adopted such an approach to be consistent with the existing snow-free SMAP official retrievals. To that end, $\mu$ was set to 20 for all pixels below $50^\degree$\,N and then is linearly reduced to zero towards the poles as the quality of VOD climatology values expectedly declines. It is worth noting that the VOD climatology is constructed using the 10-day averaged time series of pixel-level NDVI data, while the missing values in time are filled through pixel-wise linear interpolation in time \citep{o2018algorithm}, which can be uncertain over high latitudes with long-term snow cover (Fig.~\ref{fig:1}).  

The lower and upper bounds for the retrieved variables are taken as $\text{VOD}\in[0,\,1.5]$ and $\varepsilon_{\rm g}\in[0,\,60]$ \citep{bircher2016band}. These bonds are relatively wide to prevent physically unrealistic retrievals. Here, we did not use the minimum and maximum climatological values, as suggested by \citet{gao2020microwave} as the NDVI data are not reliable at high latitudes. Soil moisture climatology or porosity can also be used to define the bounds for soil permittivity. However, we avoided such a practice as the commonly used Mironov dielectric model \citep{mironov2009physically} for moist mineral soils can introduce additional errors in retrievals over organically-rich permafrost soils that transition between freeze and thaw conditions. 

\section{Data}
\label{sec:III}
\subsection{SMAP data}
We used the SMAP level-3 enhanced brightness temperatures \citep{chan2016enhanced, Chan2018} as input to the inversion algorithm at a nominal spatial resolution of 9~\si{km}. Ancillary data sets, available in the SMAP data, are also utilized including the effective ground temperature $T_{\rm g}$, surface roughness parameter $h$, vegetation single scattering albedo $\omega$, and land-cover types based on the International Geosphere-Biosphere Programme (IGBP) \citep{loveland2000development} using MODIS-MCD12QI product \citep{friedl2002global}. 

\subsection{ERA5 land dataset}
The a priori information about the 2-m air temperature and snowpack physical properties including density, temperature, and bottom melt flux is obtained from the ERA5 land dataset \citep{hersbach2018copernicus,munoz2021era5} at a resolution of 9~\si{km}. The 2-m air temperature is used to represent the canopy temperature as used previously by \citet{schwank2021temperature} based on {\it in situ} measurements. We employ snowpack melt flux and temperature as surrogate variables to identify dry snow. Specifically, a snowpack with a zero melt flux and a temperature below $-0.5^\circ$ C is considered to be dry.

\subsection{Tree height and AGB}
Due to the lack of {\it in situ} measurements on a global scale, evaluating the quality of VOD retrievals can be challenging -- especially over high latitudes. To assess the quality of VOD retrievals, previous studies \citep{rodriguez2018evaluation, li2021global, gao2021reappraisal} have suggested using canopy height and AGB as proxies for causal validations. In this study, we utilized a global tree height dataset \citep{simard2011mapping}, which has a resolution of 1~\si{km}. The dataset is based on lidar data collected in 2005 by the Geoscience Laser Altimeter System (GLAS) sensor aboard the Ice, Cloud, and land Elevation Satellite (ICESat). In areas without lidar coverage, relevant auxiliary data, such as elevation and MODIS tree cover estimates, were used to estimate tree heights using a machine-learning algorithm.

For our analysis, we also used the annual AGB dataset \citep{xu2021dataset}  at a spatial resolution of 10~\si{km}. This dataset is produced by combining multiple sources of data, including extensive forest inventory mostly from boreal and temperate regions, airborne laser scanning (ALS) data covering global tropical forests, and satellite observations. The satellite data include lidar inventory of global vegetation height structure and data from optical and microwave sensors, such as MODIS Reflectance data (MCD43A4 v006), land surface temperature (MOD11A2 v006), and radar imagery from SeaWinds Scatterometer on QuikSCAT (QSCAT).

\subsection{FLUXCOM dataset}

We used the gridded net ecosystem exchange (NEE) carbon flux dataset obtained from FLUXCOM \citep{jung2009towards,jung2011global} for causal validation of the retrieved parameters. The NEE explains the net carbon exchange between the atmosphere and land, which includes the uptake of $\text{CO}_2$ through photosynthesis and its release from soil and plant material. Positive and negative values of NEE denote net $\text{CO}_2$ fluxes into the atmosphere (carbon source) and land (carbon sink), respectively. Therefore, the spatial and temporal variation of NEE fluxes can control net changes in carbon stocks, in above- and below-ground vegetation layers as well as the soil organic carbon pools. During the winter, when photosynthesis is significantly reduced, respiration activities of plants and soil microorganisms dominate the NEE \citep{wohlfahrt2008seasonal,luers2014annual}. 

In this study, we employ the ``RS-METEO'' version of the FLUXCOM product, which provides global surface carbon fluxes at daily temporal and $0.5^\circ$ spatial resolution. It is derived by upscaling in-situ surface carbon flux measurements from the existing network of FLUXNET eddy covariance towers \citep{baldocchi2008breathing}. The upscaling process uses machine learning algorithms and globally available predictor variables from satellite observations and meteorological data. The satellite observations include daytime and nighttime land surface temperature (MOD11A226), land cover (MCD12Q127), the fraction of absorbed photosynthetically active radiation (fPAR) by the canopy (MOD15A228), and bidirectional reflectance distribution function (BRDF)-corrected reflectances (MCD43B429). The meteorological data is obtained from global climate-forcing data sets such as WATCH Forcing Data ERA Interim (WFDEI35), and the Global Soil Wetness Project-3 forcing (GSWP336). The global coverage and high quality of FLUXCOM data sets during the SMAP operation period \citep{tramontana2016predicting, jung2019fluxcom} makes it a suitable dataset for comparison purposes.

\section{Results and discussion}
\label{sec:IV}

\subsection{Outputs and comparison of the emission models}
\label{sec:IV.1}

\begin{figure}[t]
\centering\includegraphics[clip, trim=1.25cm 1.63cm 1.9cm 1.8cm, width=1.00\linewidth]{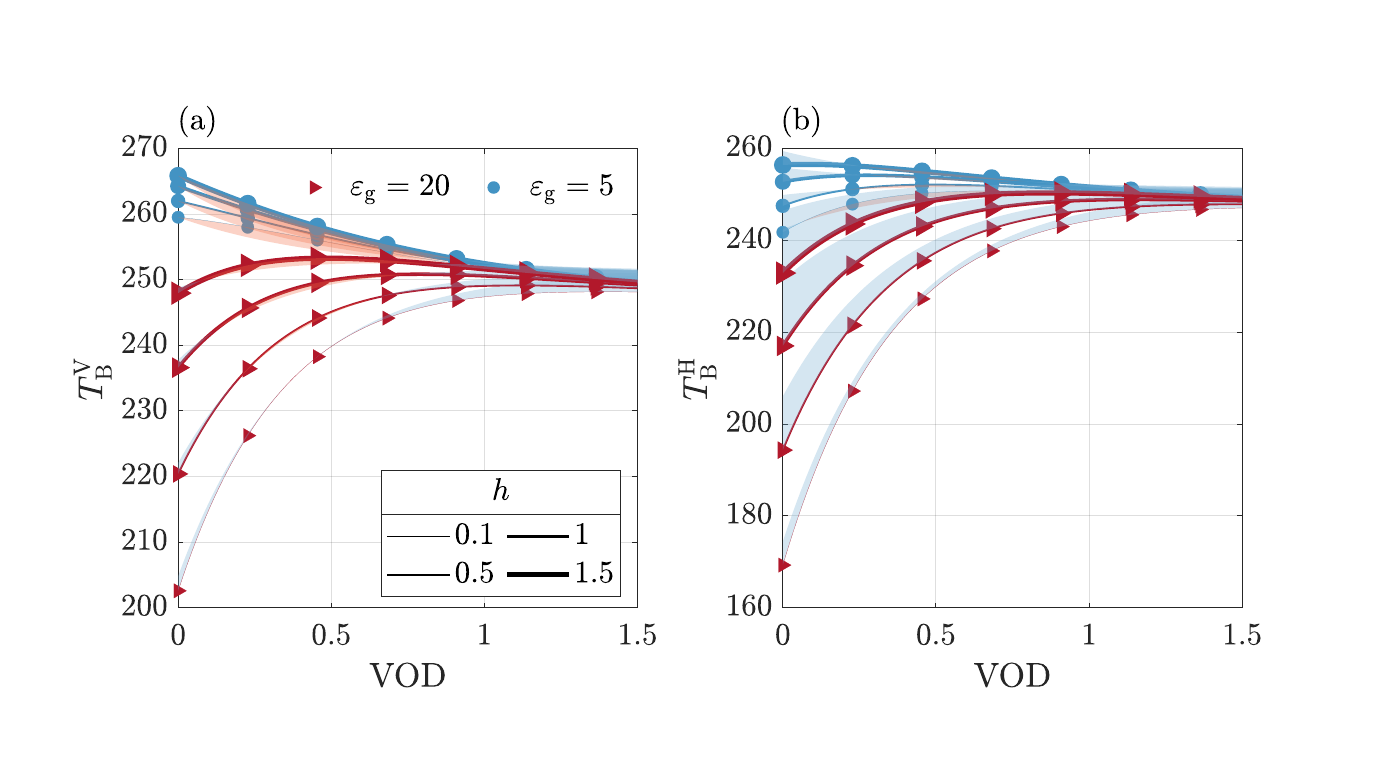}
\caption{Simulated L-band brightness temperatures as a function of VOD and roughness $h$ for the soil-snow-vegetation system with $\rho_s = 250$ at (a) vertical $T_{\rm B}^{\rm V}$ and (b) horizontal $T_{\rm B}^{\rm H}$ polarization for frozen $\varepsilon_{\rm g} = 5$ and moist $\varepsilon_{\rm g} = 20$ soils. The simulations, using the TO-snow model, are shown with solid lines while the overestimation (underestimation) of the 2-S model is depicted with a blue (red) shade.} \label{fig:3}
\end{figure}

With respect to the snowpack, the uncertainty of the retrievals of $(\tau,\varepsilon_{\rm g}$) is significantly related to the soil roughness parameter $h$ and snow density $\rho_{\rm s}$. Fig.~\ref{fig:3} shows the sensitivity of $T_{\rm B}^p$ to $h \in[0.1,\,1.5]$, for different VOD values in snow-covered surfaces both over frozen and unfrozen soils, where $T_{\rm c} = 265$~\si{K}, $\omega = 0.07$ and $\rho_{\rm s} = 250$~\si{kg.m^{-3}}. The simulations from the TO-snow model are shown by solid lines, while the differences with the 2-S model are depicted through colored shadings. To facilitate comparisons, the value of $\omega$ utilized in the TO-snow model is converted to its equivalent value in the 2-S emission model, employing the equation provided by \citet{schwank2018tau}. To simulate frozen and unfrozen ground conditions, $\varepsilon_{\rm g}$ is assumed to be 5 and 20 \citep{mironov2017temperature}, with associated $T_{\rm g} = 270$~\si{K} and $275$~\si{K}, respectively.

The results demonstrate that for moist soil, the brightness temperatures increase with increasing VOD as also shown by previous studies \citep{konings2016vegetation, entekhabi2019evaluating}. However, a reverse pattern is observed for frozen soils. This can be explained through the competing effects of canopy emission, attenuation, and soil emission on the brightness temperatures. In fact, as VOD increases, the contribution of the upwelling vegetation (soil) emission increases (decreases) in the brightness temperatures. It appears that, at SMAP observation angle, over highly emissive frozen soils, the contribution of canopy attenuation of soil emission dominates its own emission -- especially in vertical polarization. Therefore, as VOD increases, the brightness temperatures tend to decrease monotonically. However, the decrease in intensity of horizontally polarized surface emission is only apparent over very rough soils with $h\geq 1$ that exhibit higher emission intensity compared to smoother soil surfaces with $h\leq 1$.  

The brightness temperatures monotonically increase as the surface soil becomes rougher (increasing $h$) for both frozen and unfrozen moist ground conditions, which is attributed to the increase in surface soil emissivity. For the barren frozen soils ($\varepsilon_{\rm g} = 5,\tau = 0$), the horizontally polarized brightness temperature increases by less than 20~\si{K} while this warming effect exceeds 60~\si{K} for the unfrozen moist soil ($\varepsilon_{\rm g} = 20,\tau = 0$). As VOD increases, the sensitivity of the simulated brightness temperatures to $h$ decreases expectedly. For the frozen soil, the effects of roughness become almost negligible for VOD values above 1, typically corresponding to a $\text{VWC}= 10$~\si{kg.m^{-2}}. At the same time, as is shown, for rougher soil surfaces, the brightness temperatures are less sensitive to changes in VOD, and hence corresponding retrievals can be more uncertain.

The two emission models exhibit only minor differences except at horizontal polarization over the moist unfrozen ground where the TO-snow model results in colder brightness temperatures than the 2-S model by less than 15~\si{K}. This difference is more pronounced over barren ($\tau = 0$) and rougher soils. However, for $\tau<0.5$, simulated $T_{\rm B}^{\rm V}$ are nearly identical between the two models for unfrozen moist soils. For frozen soils at vertical (horizontal) polarization, the output from the 2-S model is slightly colder (warmer) than the TO-snow model, which can be due to differences between the models in the characterization of the propagation properties of snow in the L-band (Section \ref{sec:II.2}). As VOD becomes greater than 1, the TO-snow model consistently yields colder brightness temperatures than the 2-S model by less than 3~\si{K} for all soil conditions and polarization. This result is expected because the 2-S model considers internal volume scattering in the vegetation layer, while the TO-snow model does not.

\begin{figure}
\centering\includegraphics[clip, trim=1.1cm 1.6cm 1.5cm 1.8cm, width=1.00\linewidth]{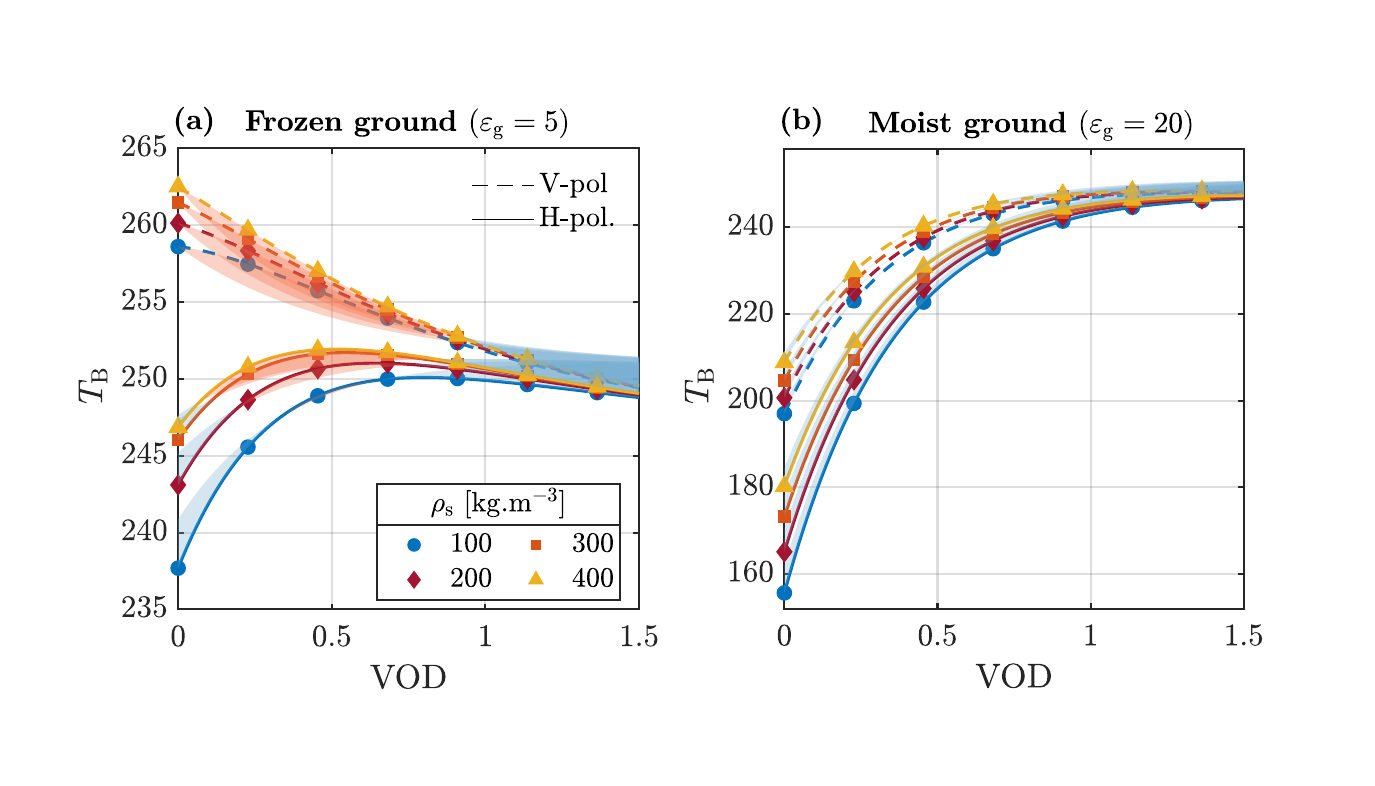}
\caption{Brightness temperatures at vertical $T_{\rm B}^{\rm V}$ and horizontal $T_{\rm B}^{\rm H}$ polarization simulated for dry snow with densities $\rho_{\rm s}=100-400$ ~\si{kg.m^{-3}} resulting from (a) frozen and (b) unfrozen moist grounds using the TO-snow and 2-S emission models. The simulations, using the TO-snow model, are shown with solid lines while the warmer (colder) differences with the 2-S model are depicted with a red (blue) shade.} \label{fig:4}
\end{figure}

Fig.~\ref{fig:4} illustrates the relationship between brightness temperatures and VOD for different values of snow density $\rho_{\rm s}$ at a fixed roughness coefficient of $h = 0.15$ while keeping all other parameters the same as used to produce Fig.~\ref{fig:3}. As expected, higher vegetation opacity leads to a monotonic increase of surface emission at both polarization channels for the moist soil. However, for the frozen ground, increased VOD decreases the surface emission at vertical polarization monotonically and increases it at the horizontal channel non-monotonically. In fact, the emission from frozen soil first increases when VOD values are less than 1 and then decreases for higher vegetation opacity due to the trade-off between canopy emission and attenuation. 

It is also apparent that the soil-snow emissivity monotonically increases as the snowpack becomes denser for both frozen and moist ground conditions. As shown, the sensitivity of brightness temperatures to $\rho_{\rm s}$ is more significant over moist soils than the frozen ones and at horizontal compared to vertical polarization in both cases. Over the barren frozen soil, the horizontally polarized brightness temperatures increase less than 10~\si{K} while this warming effect exceeds 20~\si{K} for the unfrozen moist counterpart. Moreover, for instance, when considering a typical value of $\rho_{\rm s} = 200$ and $\tau = 0.3$ over frozen ground with $\varepsilon_{\rm g} = 5$, the inclusion of snow layer results in changes of the horizontally-polarized brightness temperature by 10~\si{K}. The omission of snow, in this case, will lead to 30\% of overestimation of VOD. As VOD increases, the sensitivity of the brightness temperatures to $\rho_{\rm s}$ decreases expectedly. For both frozen and moist soils, the effects of the dry snow density on the surface emission become almost negligible for VOD greater than 1, making the retrievals of VOD independent of the presence of snow.

The shaded areas in Fig.~\ref{fig:4} show that the two emission models exhibit minor differences over vegetated surfaces with VOD smaller than 1. However, for canopies with higher VWC, over both frozen and moist, unfrozen soils, the output from TO-snow model systematically underestimates the 2-S model by less than 5~\si{K} at both polarization channels. As a result, retrievals using the TO-snow model may be prone to underestimating VOD and overestimating ground permittivity compared to the 2-S model \citep{schwank2018tau}. For VOD of less than 1, the differences between the two models are more apparent over the frozen soils and vary with the density of the snowpack. For example, for the barren frozen soil over horizontal polarization, there is an overestimation of the 2-S model by the TO-snow that shrinks from 4~\si{K} at $\rho_{\rm s} = 100$~\si{kg.m^{-3}} to 1~\si{K} at $\rho_{\rm s} = 400$\si{kg.m^{-3}}. At vertical polarization, the output from the TO-snow model is warmer up to 3~\si{K} for all snowpack density values. These deviations could be largely due to different characterizations of snow propagation properties in the two emission models as discussed in Section \ref{sec:II.2}.  

\subsection{Uncertainty quantification}
\label{sec:IV.2}

In this section, we quantify how the observation noise translates into uncertainties in retrievals of ground permittivity and VOD. For each pair of $\varepsilon_{\rm g}$ and $\tau$, brightness temperatures are simulated using the TO-snow model, considering  $\omega=0.07$, $T_{\rm g} = 273$~\si{K}, $T_{\rm c} = 265$~\si{K}, $h = 0.1$--1.5 and $\rho_{\rm s} = 100$--400~\si{kg.m^{-3}}. To account for observation noise, we perturb and generate 1000 simulated values for each pair by adding a zero-mean Gaussian random noise with a standard deviation of 1~\si{K}. The corresponding $\varepsilon_{\rm g}$ and $\tau$ values were then retrieved using Eq.~\ref{eq:06} with $\mu=0$.

\begin{figure}
\centering\includegraphics[width=1\linewidth]{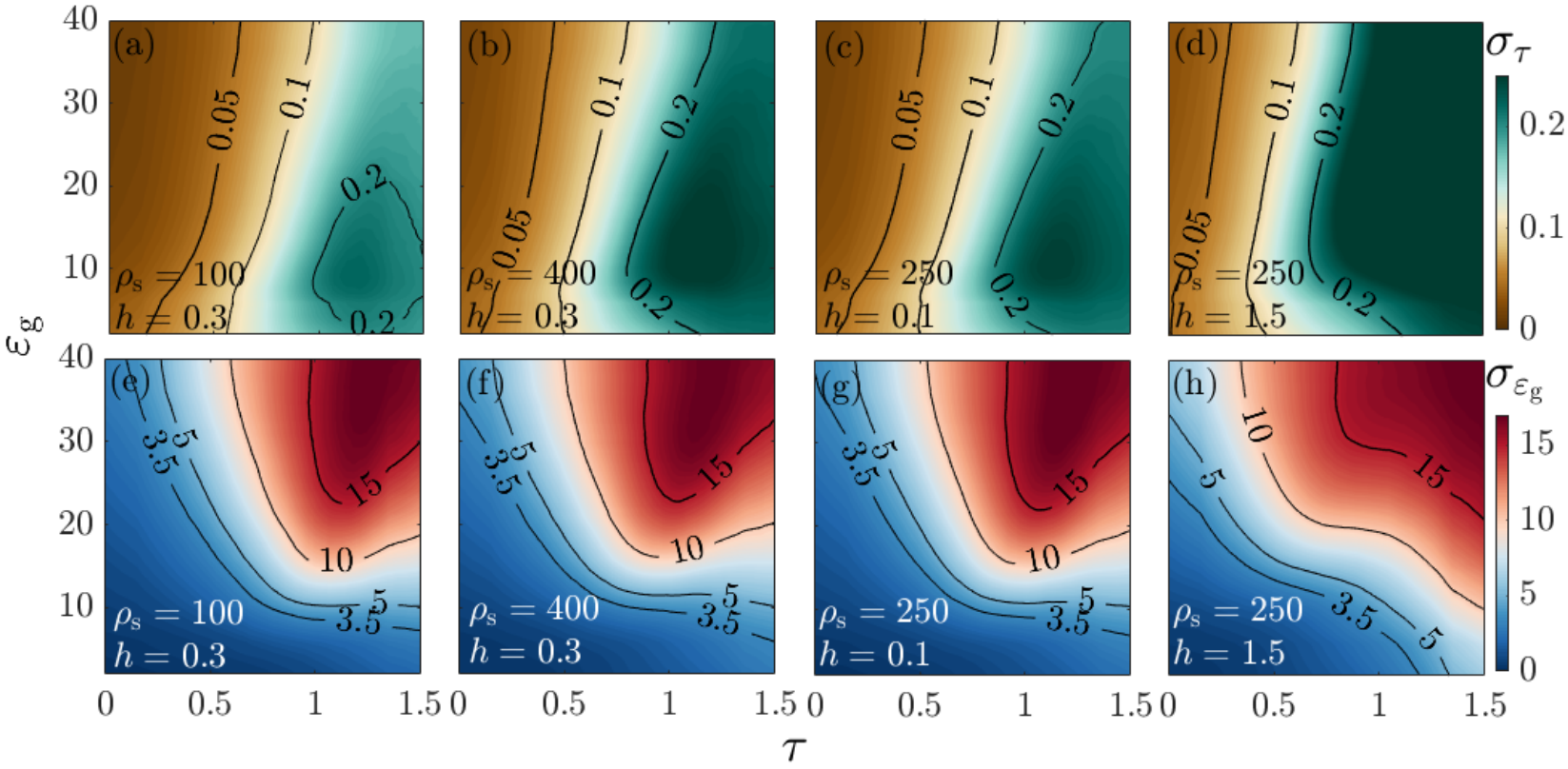}
\caption{Isolines of the error standard deviation in the retrievals of (a-d) VOD $\tau$ and (e-h) ground permittivity $\varepsilon_{\rm g}$, in the presence of a zero-mean Gaussian observation noise with a standard deviation of 1~\si{K}.} \label{fig:5}
\end{figure}

Fig.~\ref{fig:5} shows the contour lines of the standard deviation of the retrieval error for $\tau$ ($\sigma_\tau$, first row) and $\varepsilon_{\rm g}$ ($\sigma_{\varepsilon_{\rm g}}$, second row), respectively, for varying values of the roughness coefficient and snowpack density. The first two panels in each row characterize the error for a range of snow density and a mean roughness value of 0.3, while the last two panels in each row show the effects of soil roughness 0.1--1.5 for $\rho_{\rm s}=$250~\si{kg.m^{-3}}. As shown in Fig.~\ref{fig:5}\,a--d, $\sigma_\tau$ is heteroscedastic with respect to VOD but does not change significantly as a function of ground permittivity.  This means that the error increases as the mean VOD increases and exhibits a higher value in the presence of high VWC especially when the soil is dry or frozen. This can be attributed to the reduced sensitivity of brightness temperatures to changes in high VOD values of greater than 1 over highly emissive soils.

Fig.~\ref{fig:5}\,a,b shows that this heteroscedastic pattern can be a function of $\rho_{\rm s}$. For instance, over moist soils with $\varepsilon_{\rm g} = 20$ and $\tau = 0.5$, $\sigma_\tau$ increases by 45\%, from 0.06 to 0.087, as the $\rho_{\rm s}$ changes from 100 to 400~\si{kg.m^{-3}}. This increase can be largely related to the reduction in the effective reflectivity of the soil-snow system that leads to reduced sensitivity of brightness temperatures to changes in VOD, as the snow becomes denser. As evidenced, increasing roughness can have a more significant impact on the changes of $\sigma_\tau$ than the snow density (Fig.~\ref{fig:5}\,c,d). The results show that the contour lines of $\sigma_\tau=0.2$ (dark green areas) expand towards lower ranges of VOD as the roughness increases. In fact for $\tau = 0.5$ over moist soils with $\varepsilon_{\rm g} = 20$, the error almost doubles, from 0.07 to 0.13, as roughness increases from 0.1 to 1.5. This observation suggests that soil roughness has a much more significant impact on reduced sensitivity of the brightness temperatures to VOD than the snow density -- as discussed in Section~\ref{sec:IV.1}. 

The results in Fig.~\ref{fig:5}\,e--h indicate that $\sigma_{\varepsilon_{\rm g}}$ exhibits larger uncertainties as both the soil becomes wetter and the VWC increases. Therefore, unlike the previous case, the magnitude of error is not only a function of VOD but also soil moisture. Note that the error of $\varepsilon_{\rm g}$ seems to be almost independent of snowpack density for a constant roughness (Fig.~\ref{fig:5}\,e,f); however, when the roughness varies from 0.1 to 1.5, the error pattern changes significantly (Fig.~\ref{fig:5}\,g,h). For instance, the contour lines $\sigma_{\varepsilon_{\rm g}}=5$, propagate towards lower values of ground permittivity and VOD and increase by almost 50\% from 5 to 8 in the case of a moist ground $\varepsilon_{\rm g} = 20$ with $\tau = 0.5$  -- as the roughness parameter increases from 0.1 to 1.5.

Overall, the analysis indicates that $\sigma_\tau$ is smaller than 0.1 for VOD values of less than 0.5 (i.e., VWC$\approx$ 5~\si{kg.m^{-2}}), snow densities 100--400~\si{kg.m^{-3}} and roughness coefficients 0.1--1.5. A simple Monte-Carlo simulation, using the Mironov soil dielectric model \citep{mironov2009physically}, shows that the SMAP target soil moisture error standard deviation of 0.04~\si{m^3.m^{-3}} translates into a standard deviation of 3.5 in the space of ground permittivity, Therefore, in Fig.~\ref{fig:5}\,e--h, the area below the contour line 3.5 denotes the acceptable range of the retrieval parameters -- considering that soil is moist below the snowpack.                      

\subsection{Causal validation}

As previously noted, validation of VOD and ground permittivity in the presence of snow cover is challenging due to the scarcity of suitable {\it in situ} measurements. However, it is a common practice to causally validate these retrievals by comparing them with the space-time dynamics of naturally correlated variables such as NEE \citep{teubner2018assessing,hunt1996global}, the temperature of air and ground, and with other vegetation-related proxies such as tree height and AGB \citep{rodriguez2018evaluation, li2021global, gao2021reappraisal}. To conduct this analysis, we first examine the performance of the emission model over a daily SMAP orbital retrieval and then expand it for five years (2015--2020) of retrievals during the winter months from October to April. 

\subsubsection{Orbital retrievals}
\label{sec:IV.3}

\begin{figure}[t]
\centering\includegraphics[width=1\linewidth]{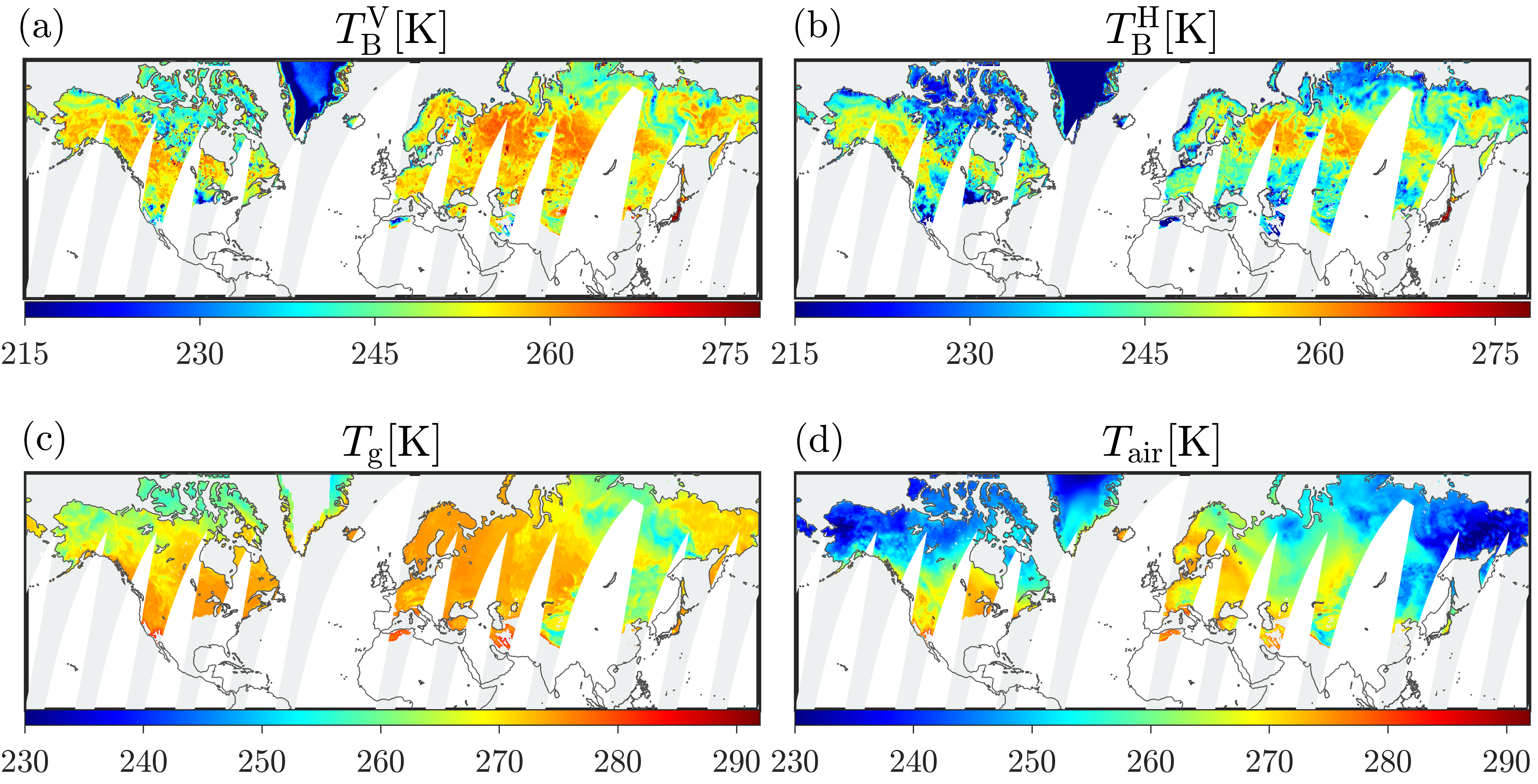}
\centering\includegraphics[width=1\linewidth]{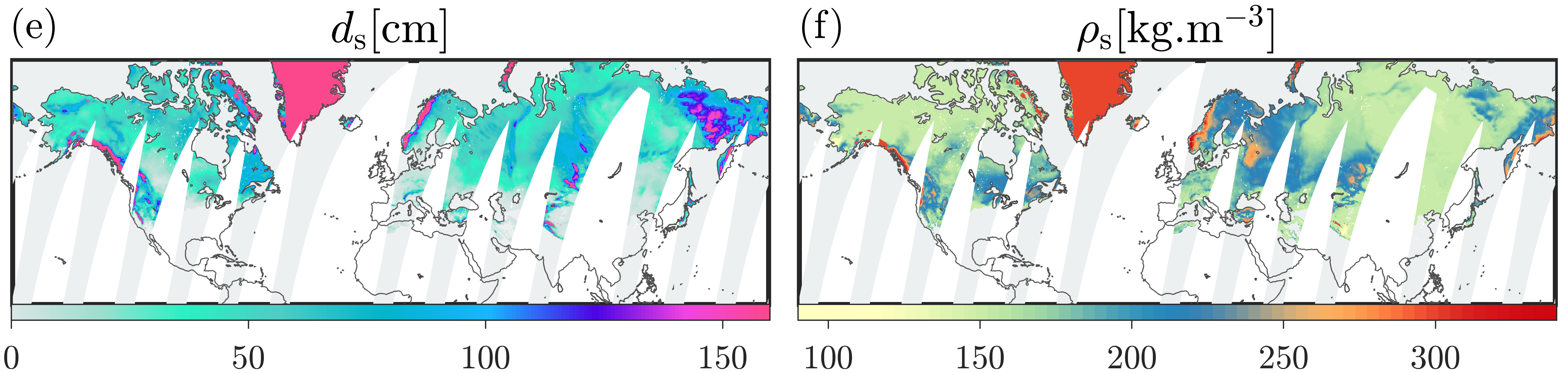}
\caption{The SMAP level-III enhanced brightness temperatures on January 23, 2017, at (a) vertical and (b) horizontal polarization, (c) the effective ground temperature, as well as ERA5 (d) 2-m air temperature, (e) snowpack depth, and (f) density.} \label{fig:6}
\end{figure}

Fig.~\ref{fig:6}\,a--d shows the observed $T_{\rm B}^p$, as well as ground and air temperatures mapped onto SMAP orbits on January 23, 2017. The data are shown only over snow-covered areas for which the snow depth and density are displayed in Fig.~\ref{fig:6}\,e,f. 

The observed brightness temperatures show noticeable variability and cold depressions across different regions of the world. In North America, these depressions can be observed over the Arctic Archipelago, north of Alaska, and the western flanks of the Sierra Nevada Mountains that extend up to Southern Washington and Western Idaho. Over the Midwest United States, the cold depressions below 230~\si{K} extend from southern Minnesota to northern Iowa and Illinois. Similarly, in Europe, radiometrically cold areas can be seen to expand from Norway and the southern parts of Sweden to Switzerland, covering Germany and France. In Asia, the orbital tracks passing over the Northern Siberian Plateau, Kazakhstan, Mongolia, Tibetan Plateau, and northeastern China, exhibit significantly lower brightness temperatures than the surrounding areas, with more pronounced depressions being observed in horizontal compared to vertical polarization.

These radiometrically cold regions are often covered with a thick $d_{\rm s}>50$ ~\si{cm} and dense $\rho_{\rm s}>200$ ~\si{kg.m^{-3}} insulating snow cover (Fig.~\ref{fig:6}\,e,f). Therefore, the observations indicate the likelihood of the presence of moist soils below the snowpack. The reanalysis data indicate that the ground temperatures largely vary between 273--275~\si{K} (Fig.~\ref{fig:6}\,c), despite having the air temperature far below the freezing point (Fig.~\ref{fig:6}\,d). On the other hand, in southern Alaska, northwest Canada, and central Russia there exist regions with warmer brightness temperatures than their surrounding areas due to highly emissive frozen grounds with a temperature of around 260--270~\si{K}. It appears that in these regions, the presence of vegetation and snowpack with mean $\rho_{\rm s} = 180- 240$~\si{kg.m^{-3}} may exacerbate the emission signal, for example, over the boreal forests in western Russia.

\begin{figure}[t!]
\centering\includegraphics[width=0.65\linewidth]{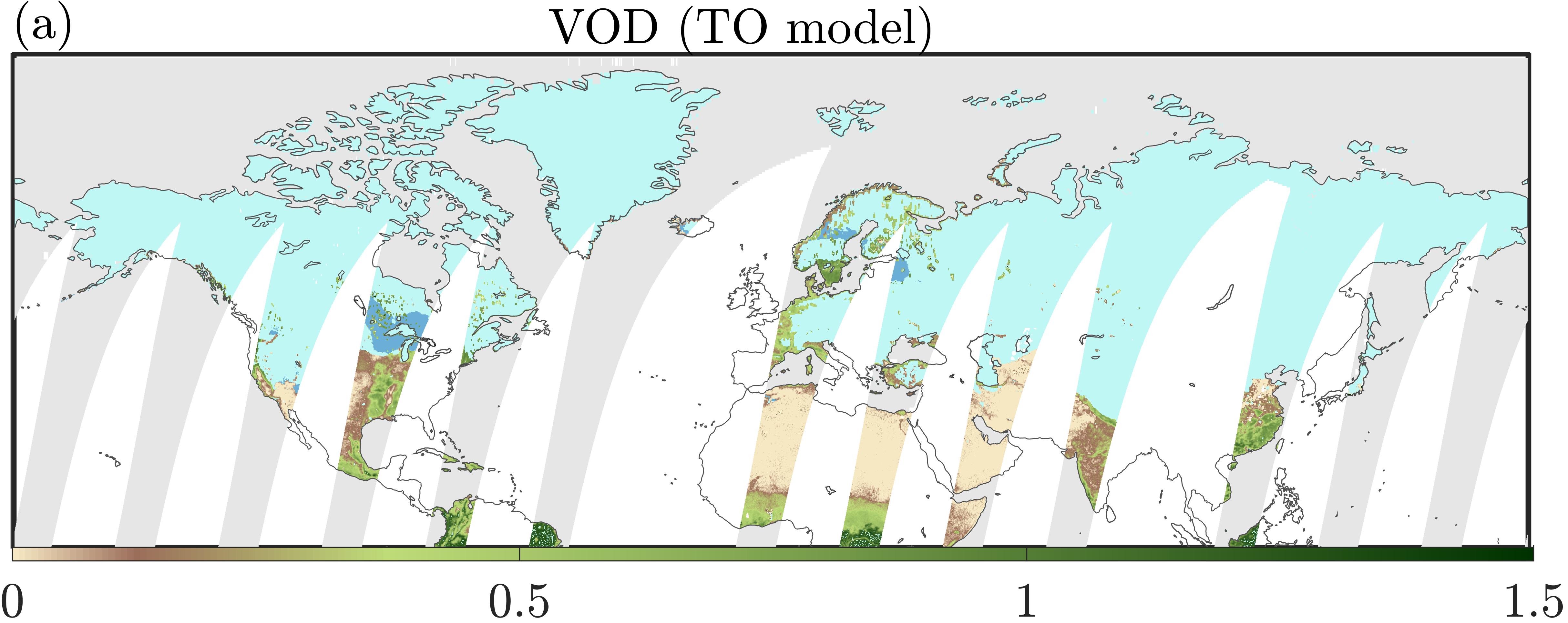}
\centering\includegraphics[width=0.65\linewidth]{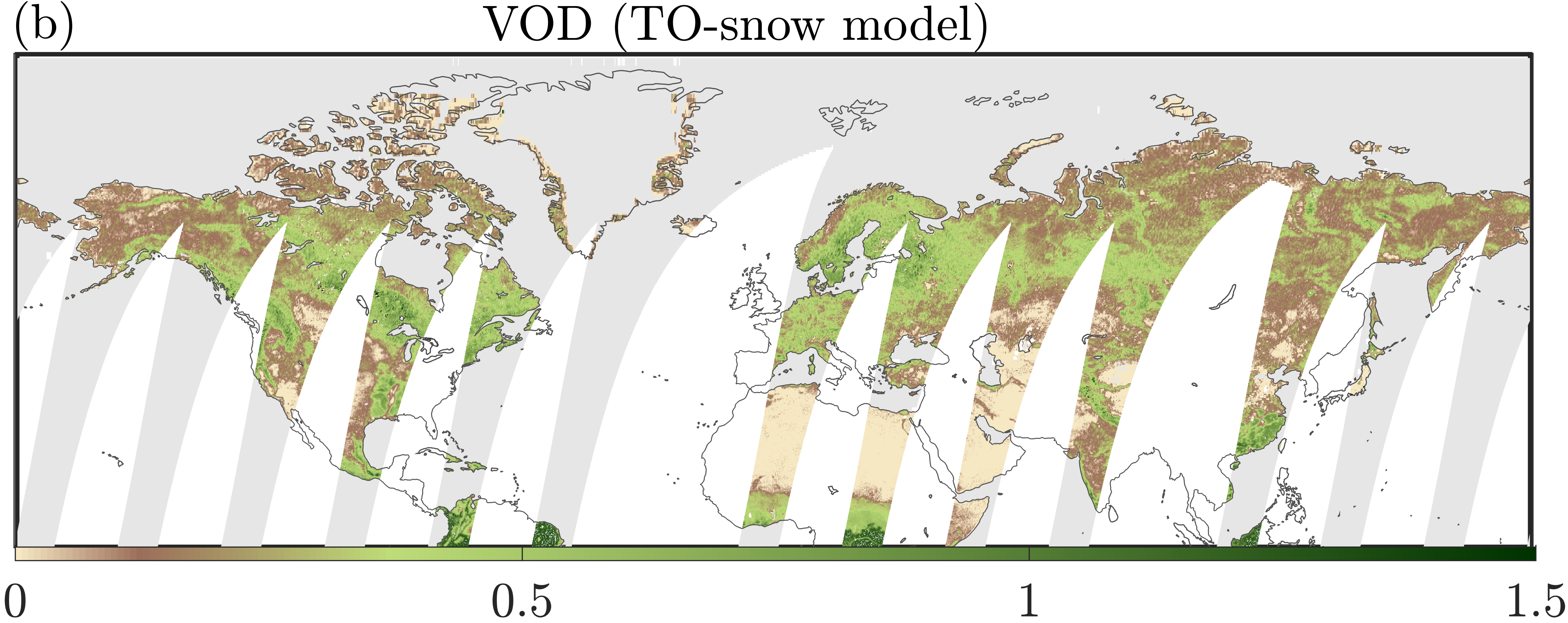}
\centering\includegraphics[width=0.65\linewidth]{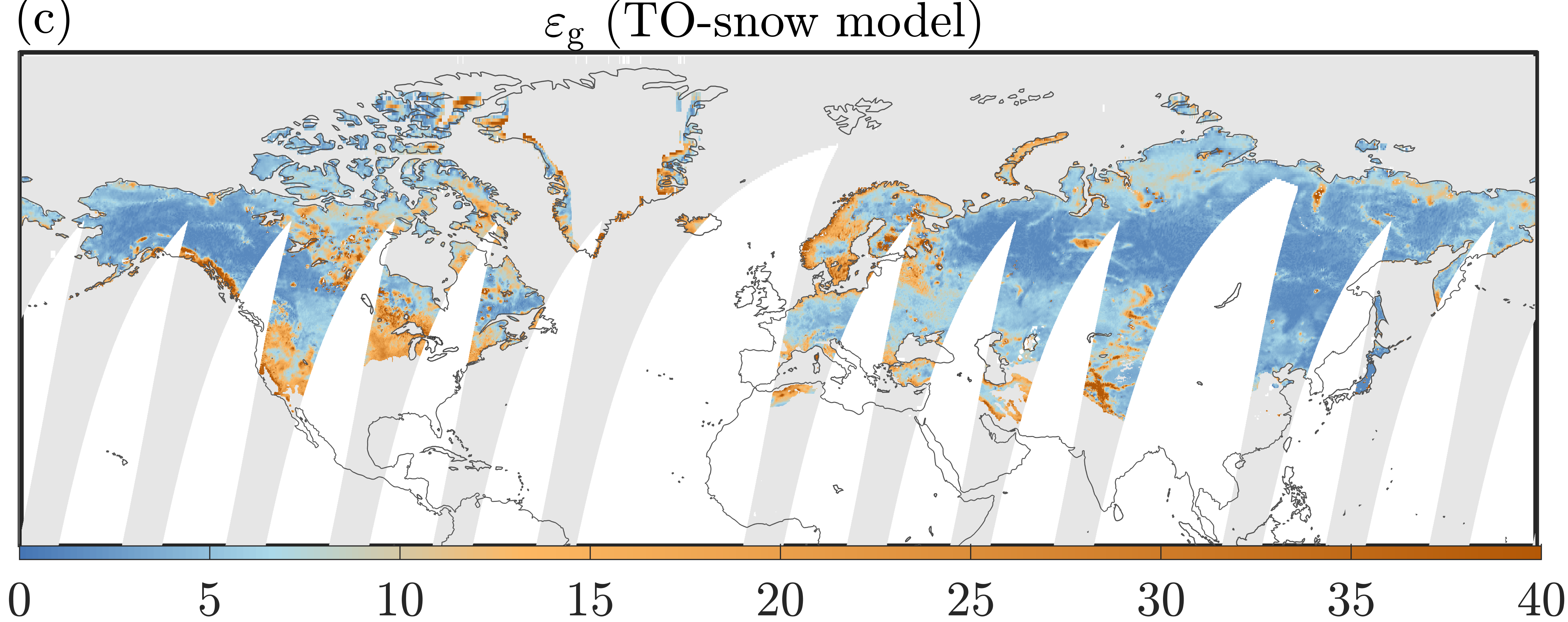}
\caption{(a) Official SMAP VOD product using the dual channel algorithm on January 23, 2017, in which blue areas represent pixels where VOD was not retrievable due to the presence of snow cover, and retrieved (b) VOD and (c) ground permittivity below snowpack through inversion of the TO-snow model. Darker shaded blue areas in the top panel represent the pixels that are flagged as potential wet snow -- identified as explained in the text.} \label{fig:7}
\end{figure}

Fig.~\ref{fig:7}\,a displays the official SMAP product of VOD retrievals obtained using the dual-channel algorithm \citep{chaubell2021regularized}. The areas covered with potentially dry (wet) snow are depicted in light (dark) blue shading, where SMAP official retrievals are not available. It should be noted that the presence of wet snow affects the emission signal as the penetration depth in L-band decreases as snow wetness increases. Consequently, above 5\% to 10\% of snow liquid water content \citep{matzler1984microwave}, the emission signal can be saturated, and the observed emissivity becomes predominantly a function of snow wetness and VWC. Additionally, the relationship between L-band brightness temperatures and snow wetness is not necessarily monotonic \citep{naderpour2017davos, naderpour2022band}. Therefore, the presented retrievals under wet snow conditions should be considered uncertain. 


Fig.~\ref{fig:7}b,c shows retrievals of VOD and ground permittivity, respectively using the TO-snow model. We need to emphasize that, we kept the official SMAP retrievals and just filled the retrieval gaps where snow cover is present. Overall, visual inspection shows that the spatial pattern of the retrievals is consistent with the observed brightness temperatures, and no abrupt changes in VOD are observed at the boundaries of snow-covered surfaces. For instance, over the Russian and West Siberian Plains warm signatures in horizontal (vertical) brightness temperatures are observed varying from 255 to 260 (260 to 265)~\si{K}. In these regions, retrievals show high VOD values $\tau>0.5$ and low ground permittivity ranging between 3 and 5. These observations suggest the presence of relatively high VWC in the canopy with almost frozen grounds at the surface.

Another example is the cold signatures in vertically (horizontally) polarized brightness temperatures with values smaller than 240 (230)~\si{K} over the United States in Washington, Oregon, Idaho, and Nevada, as well as coastal areas in Canada's western province of British Columbia, and Norway. Over these regions, the retrievals show high values of $\varepsilon_{\rm g}>20$, indicating the presence of unfrozen moist grounds below dense ($\rho_{\rm s}>250$~\si{kg.m^{-3}}) and thick ($d_{\rm s}>70$~\si{cm}) snowpack. 

\begin{figure}

\centering\includegraphics[width=1\linewidth]{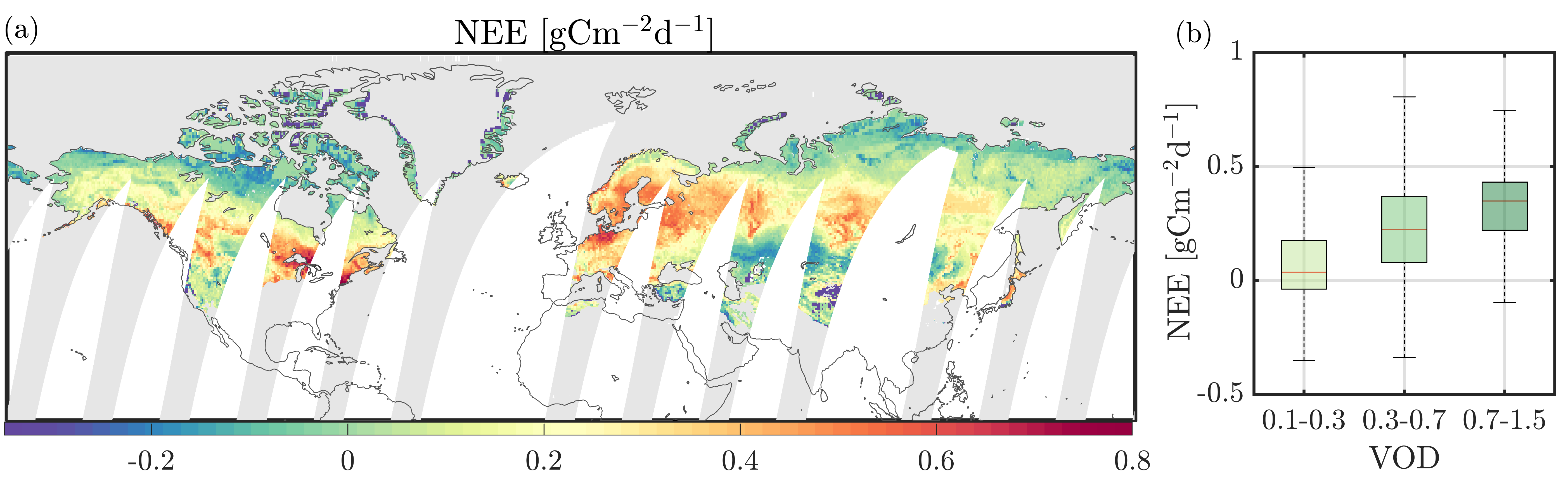}
\caption{(a) Spatial distribution of daily net ecosystem exchange (NEE) on January 23, 2017, obtained from FLUXCOM and (b) its statistics for three ranges of retrieved VOD values shown in Fig.~\ref{fig:7}\,b. In the box-whisker plot, the boxes show the $25^{th}$ and $75^{th}$ percentiles around the median and the whiskers extend to 1.5 times the interquartile range.} \label{fig:8}
\end{figure}

In addition, we evaluate the performance of the orbital VOD retrieval by comparing it with NEE carbon flux data. Previous studies have demonstrated a strong positive dependence between VOD and NEE \citep{teubner2018assessing}, with NEE effectively used to assess interannual carbon dynamics of vegetated land surfaces on a global scale \citep{dou2023reliability}. This correlation has been leveraged to generate carbon flux data sets using microwave remote sensing observations of VOD at Ku, X, and C-band on a global scale from 1988 to 2020 \citep{wild2022vodca2gpp}. Since respiration is the primary contributor to NEE during winter months, we expect to observe higher NEE values over organic soils covered with relatively dense vegetation.

Fig.~\ref{fig:8}\,a shows daily NEE fluxes mapped onto the SMAP orbit on January 23, 2017. Visual inspection reveals a significant positive spatial correlation between the retrieved VOD (Fig.~\ref{fig:7}\,b) and NEE. For example, over the boreal forest of central Canada, Europe, and eastern Russia, where VOD exceeds 0.5, NEE values greater than 0.5~\si{g.C.m^{-2}d^{-1}} are observed. In contrast, areas with a VOD of less than 0.2, such as the western United States, northern Russia, Kazakhstan, and northwest China, exhibit negligible NEE values. To quantify the dependency, we stratified NEE fluxes based on low, moderate, and high ranges of the retrieved VOD (Fig.~\ref{fig:8}\,b). The results indicate that when VOD decreases from high to low, the median value of NEE decreases by nearly 85\% from 0.4 to 0.05~\si{g.C.m^{-2}d^{-1}}.

\subsubsection{Annual retrievals}

\begin{figure}[t!]
\centering\includegraphics[width=0.8\linewidth]{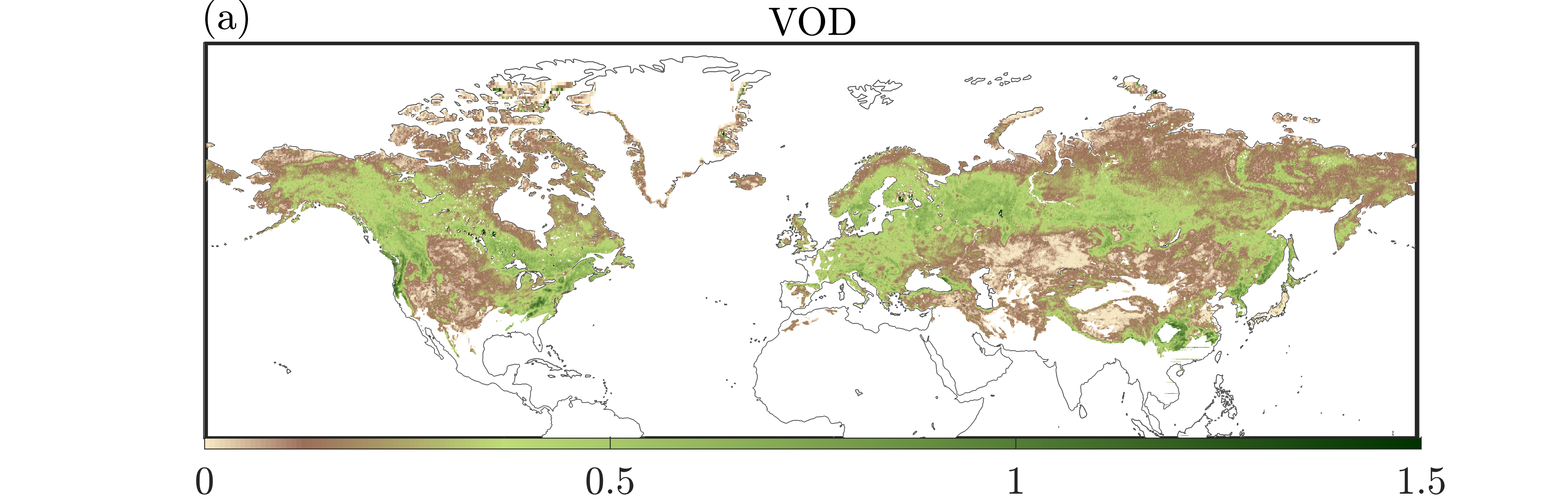}
\centering\includegraphics[width=0.8\linewidth]{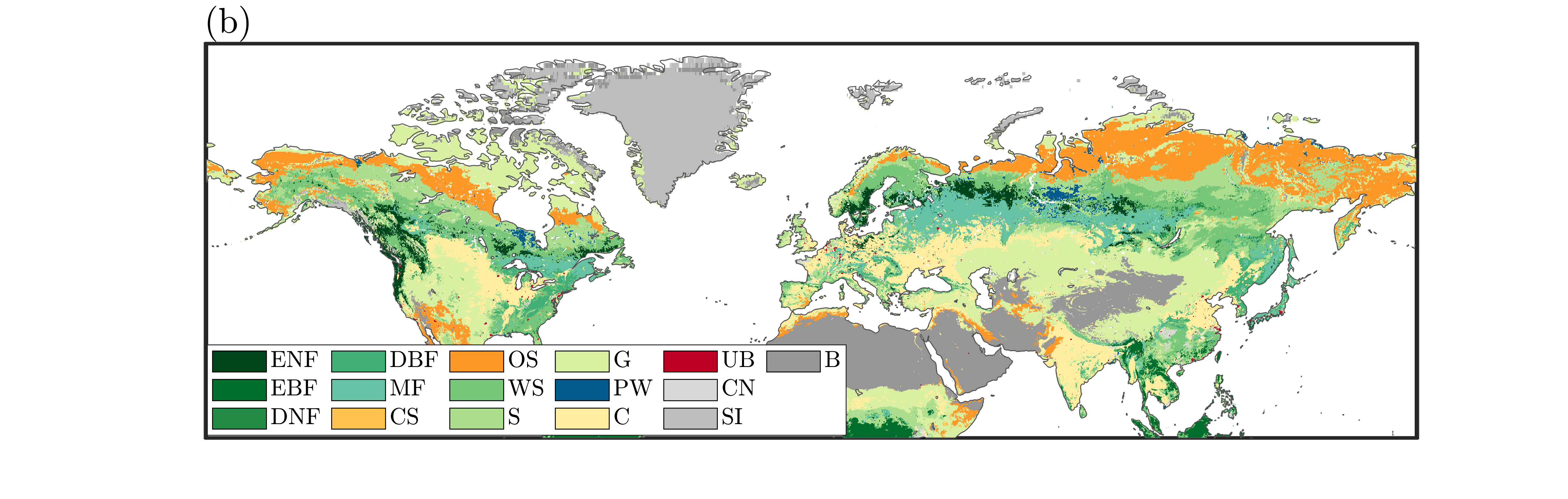}
\centering\includegraphics[width=0.6\linewidth]{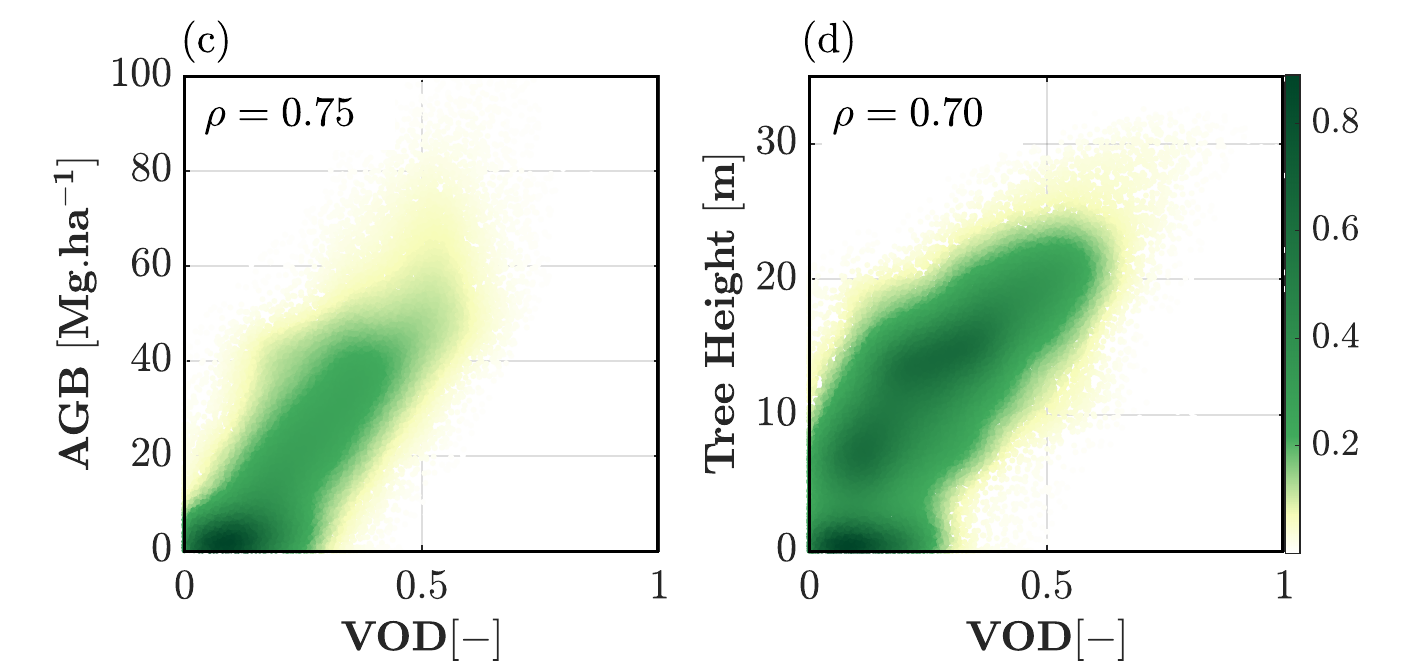}
\caption{(a) Mean values of the retrieved VOD from October to April 2015--2020, (b) IGBP land cover map obtained from MODIS Land Cover Dynamics (MCD12C1), and density scatter plots of the mean VOD estimates versus (c) AGB and (d) tree height -- where the color map represents a measure of the density of data points. The land cover map includes evergreen needle-leaf forests (ENF), evergreen broadleaf forests (EBF), deciduous needle-leaf forests (DNF), deciduous broadleaf forests (DBF), mixed forests (MF), closed shrublands (CS), open shrublands (OS), woody savannas (WS), savannas (S), grasslands (G), permanent wetlands (PW), croplands (C), urban and built-up lands (UB), croplands/natural vegetation mosaics (CN), snow and ice (SI) and barren (B).} \label{fig:9}
\end{figure}

This section presents global maps of time-averaged retrievals for VOD and ground permittivity over snow-covered areas. The spatial variability of the mean retrievals is causally validated with respect to land cover types, vegetation proxies, and the probability of ground temperature exceeding the freezing point of water. The retrievals were obtained by averaging over a period of five years, from 2015 to 2020, during the winter months from October to April. Due to a potentially high level of uncertainty, wet snow pixels were excluded from the analysis. Fig.\ref{fig:9}\,a,b show that the spatial pattern of mean VOD is consistent with the land cover types. As shown, mean VOD values greater than 0.5 are over areas covered with woody vegetation and savannas. In contrast, areas with short and sparse vegetation such as shrublands, grasslands, and croplands, exhibit significantly lower VOD values of less than 0.2.

We observe VOD values greater than 0.5 over Scandinavia, Russian Taiga, central Siberia, and the Canadian boreal forest, where land cover type is dominated by evergreens, mixed forests, and savannas that contain coniferous plant species \citep{shorohova2009natural}. Similarly, in deciduous and mixed forests extending from the Appalachian Mountains to the national forests in the Pacific Northwest, as well as in Eastern Asia from northeast China to southeast Russia, VOD values greater than 0.5 are observed, despite the presence of leafless biomes during the winter. This can be attributed to the correlation of VOD with vegetation water content, which is not synchronous with leaf development in deciduous forests, particularly during the winter \citep{tian2018coupling}. However, areas covered with shrublands and grasslands in Canadian Tundra and Palearctic Tundra in Eurasia are typically submerged by snow \citep{eastman2013global} during winter and therefore exhibit significantly lower VOD values, usually below 0.1.

Fig.~\ref{fig:9}\,c,d shows the spatial dependencies between the mean VOD values and vegetation proxies including tree height and AGB. We observe a strong correlation with coefficients 0.75 and 0.70, respectively. This dependency is almost linear for AGB but slightly nonlinear for tree height. It is worth noting that the found correlations are consistent with the earlier results reported in \citep{rodriguez2018evaluation, gao2021reappraisal, li2021global}.

\begin{figure}[t!]
\centering\includegraphics[width=0.8\linewidth]{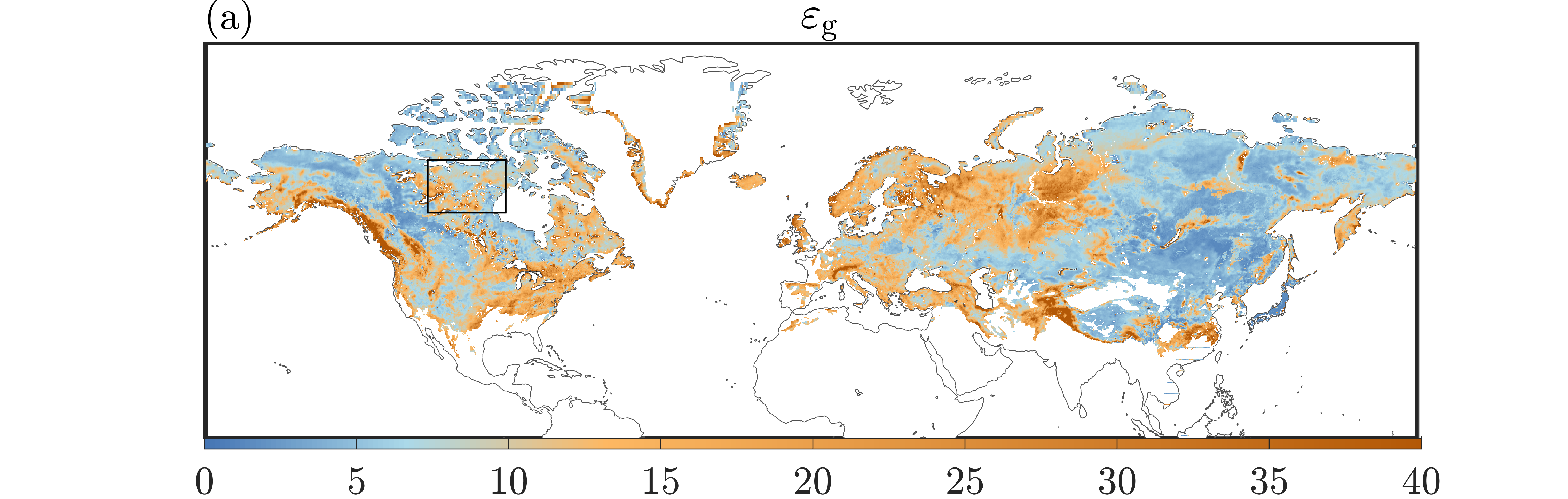}
\centering\includegraphics[width=0.8\linewidth]{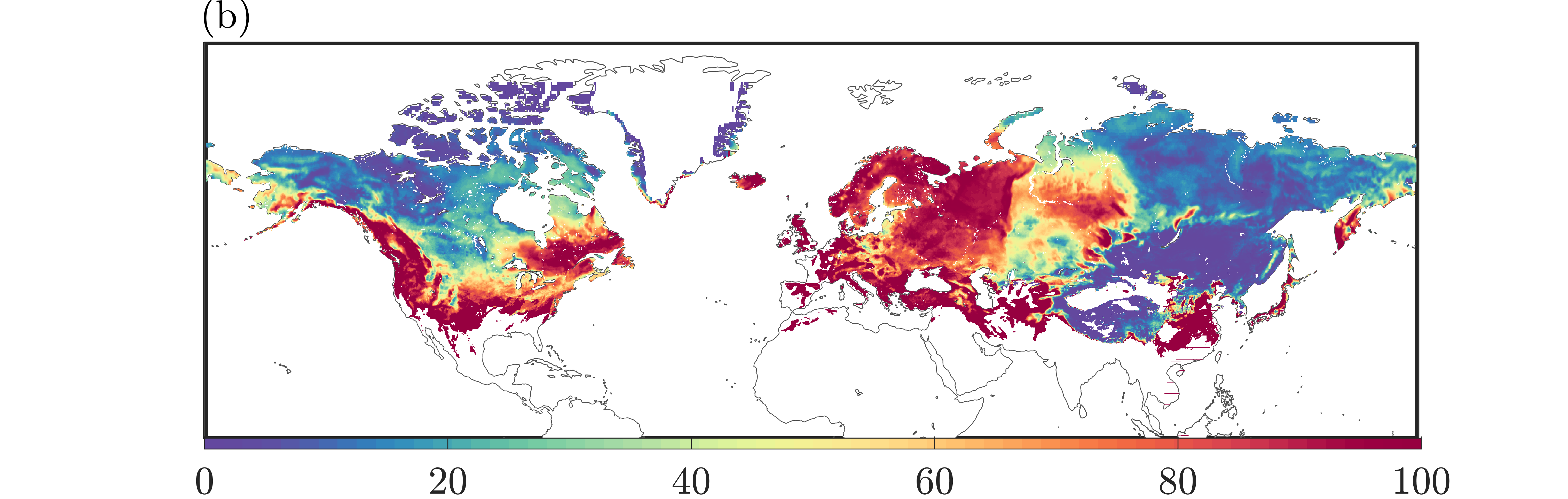}

\caption{(a) Mean values of retrieved ground permittivity from October to April during 2015--2020, and (b) corresponding probabilities of ground temperature exceeding the freezing point of water during the same period.} \label{fig:10}
\end{figure}

The mean retrieved ground permittivity values and the probabilities of ground temperature exceeding the freezing point of water are shown in Fig.~\ref{fig:10}. As is evident, regions with an exceedance probability of 70\% have a mean ground permittivity greater than 15. These regions are primarily in North America below $45^\circ$ N, especially over forested areas on the East and West Coasts, as well as in Eurasia extending from Europe to western Siberia in Russia. In contrast, areas such as northern Alaska, Northwest Territories in Canada, eastern Siberia, and Far East Russia exhibit lower permittivity values of less than 6, with an unfrozen ground probability of less than 10\%, indicating a high likelihood of frozen soils beneath the snowpack for a longer duration in the winter.

It is interesting to note that several areas exhibit relatively high ground permittivity values ranging between 10 and 15, despite having low exceeding probabilities of less than 30\%, for example over the Nunavut territory in Canada shown by a black bounding box in Fig.~\ref{fig:10}\,a. This can be attributed to the abundance of large water bodies including lakes and wetlands that are deep enough and may not be fully frozen during the winter. At the same time, the role of organically rich permafrost soils that can retain more unfrozen water below the freezing point, shall not be overlooked \citep{mavrovic2021soil}. 

\subsubsection{Impacts of dry snow on VOD retrievals}

\begin{figure}[t]
\centering\includegraphics[width=0.8\linewidth]{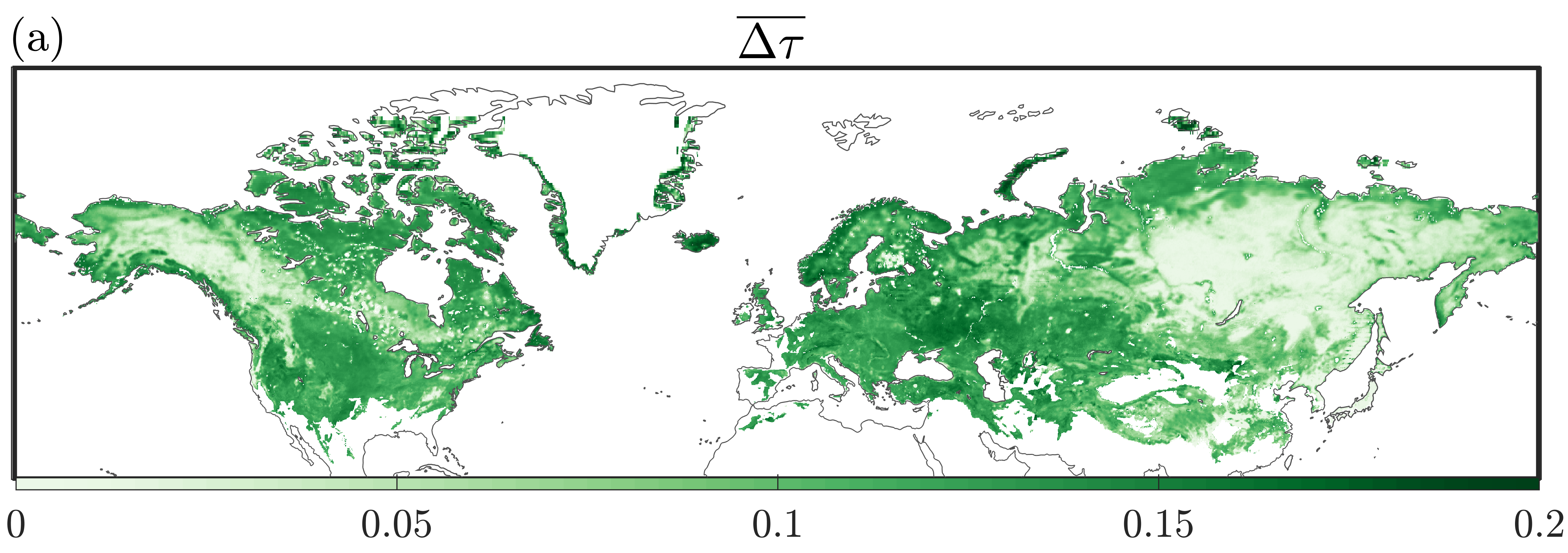}
\centering\includegraphics[width=0.8\linewidth]{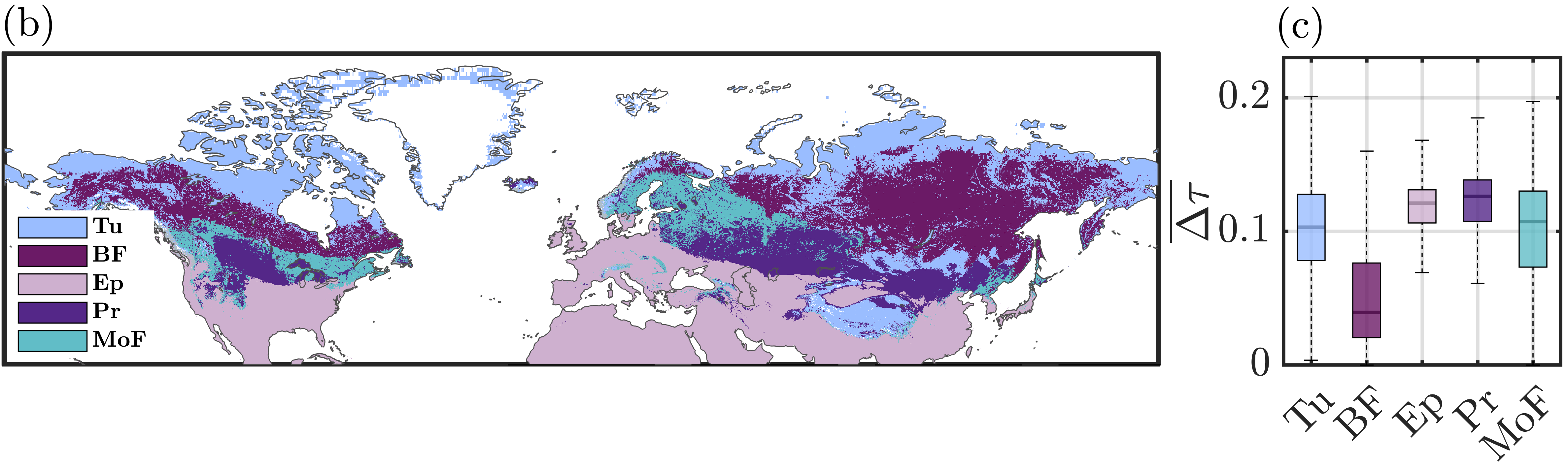}

\caption{(a) The biases in VOD retrievals, obtained by considering snow effects versus retrievals obtained by omitting the effects of snow cover (setting $\rho_{\rm s} = 0$) in the TO-snow emission model, (b) Dominant snow cover types \citep{sturm2021revisiting} including tundra (Tu), boreal forest (BF), ephemeral (Ep), prairie (Pr), and montane forest (MoF), and (c) the biases stratified according to snow cover types. In the box–whisker plot, the boxes show the 25th and 75th percentiles around the median and the whiskers extend to 1.5 times the interquartile range.} \label{fig:11}
\end{figure}

One of the key questions we aim to answer here is -- what is the spatial distribution of VOD overestimation when the presence of (dry) snow cover is ignored? In order to answer this question, focusing on the Northern Hemisphere, we retrieve the VOD values for all SMAP overpasses from 2015 to 2020 with ($\tau$) and without considering the presence of snow ($\tilde{\tau}$). The annual value of $\overline{\Delta \tau} = \overline{\tilde{\tau}-\tau}$ is shown in Fig.~\ref{fig:11}\,a for winter months from October to April. The statistics of the difference for major snow-cover classes (Fig.~\ref{fig:11}\,b) are shown in Fig.~\ref{fig:11}\,c. 

As anticipated, the VOD retrievals exhibit a clear overestimation when the effects of snow cover are not properly considered, and it becomes more pronounced with increasing snowpack density. The boreal forest (BF) snow cover type, found in Eastern Russia and northwest North America, exhibits minimum differences with a median value of 0.05 (i.e., VWC$\approx$0.5 \si{kg.m^{-2}}). This is attributed to a relatively light snow cover of this type, with an average density of approximately 150~\si{kg.m^{-3}}. In contrast, the maximum differences, with a median value of approximately 0.15, are observed over the ephemeral and prairie snow cover types, where densities can go up to 400~\si{kg.m^{-3}} during the spring. As is evident, the widest uncertainty bound, with a 95\% confidence bound of 0.2, occurs over the tundra which typically consists of wind-slab over depth-hoar and montane forest snow-cover types with highly variable density, ranging from 100 to 400~\si{kg.m^{-3}} \citep{sturm2021revisiting}.  

These biases are also influenced by ground conditions, as previously demonstrated in Fig.~\ref{fig:4}. In fact, the warming effects of the snow density are more pronounced when the soil below the snowpack is wet, which can lead to a larger VOD overestimation. This is particularly apparent in regions extending from Europe to western Siberia in Russia, where the soil temperature remains above the freezing point of water for approximately 70\% of the time during winter (Fig.~\ref{fig:10}\,b). As shown, over these regions, the overestimation of VOD can exceed 0.15. Conversely, regions with boreal forest snow cover over northern Alaska, eastern Siberia, and Far East Russia experience frozen ground conditions for more than 90\% of the winter, which is aligned with the observed lower biases.

\subsubsection{Time-series analysis}

\begin{figure}[t!]
    \centering
    \includegraphics[width=1\linewidth, trim = 0cm 2.8cm 0cm 2.8cm]{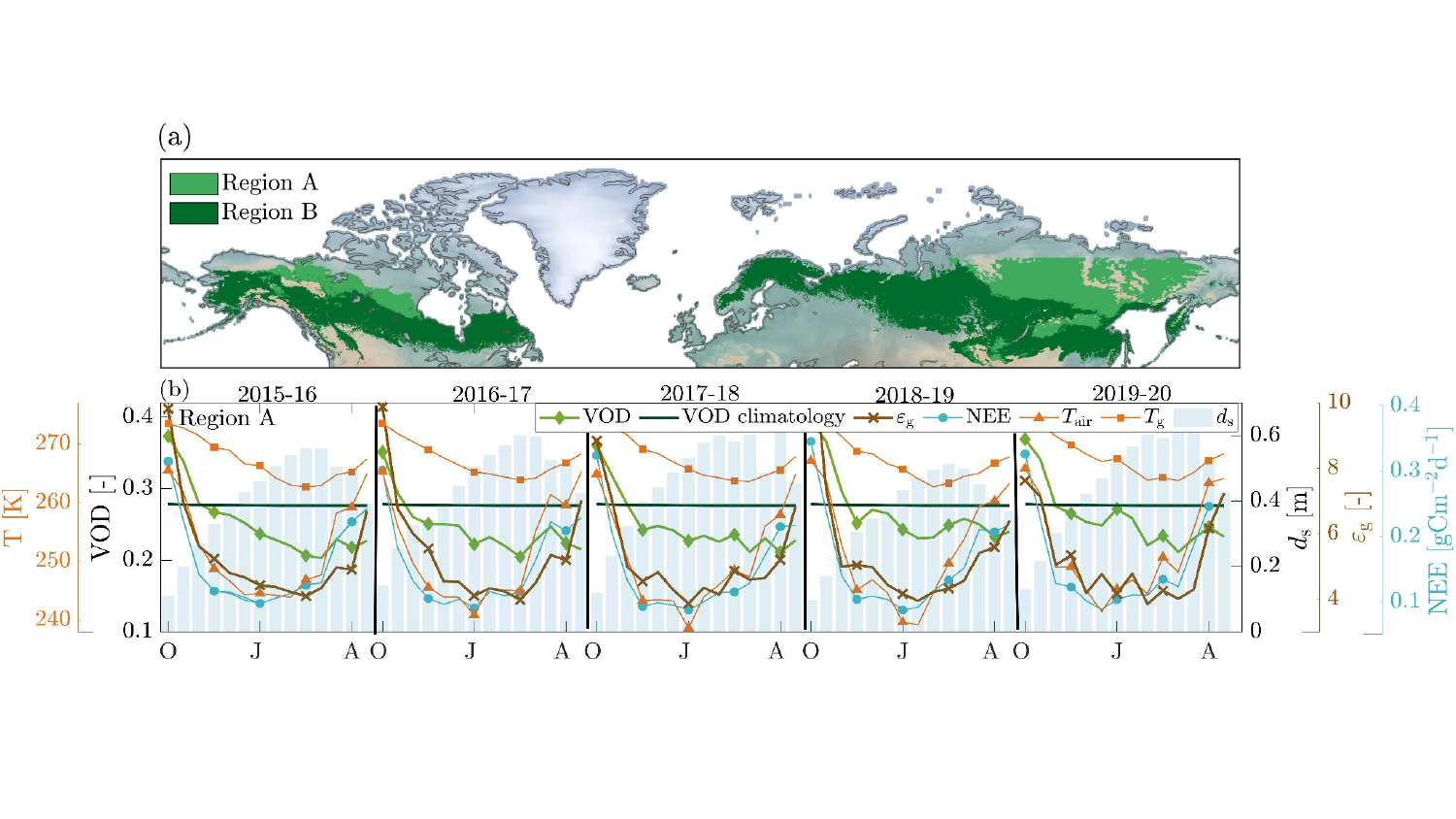}\\
  \includegraphics[width=1\linewidth]{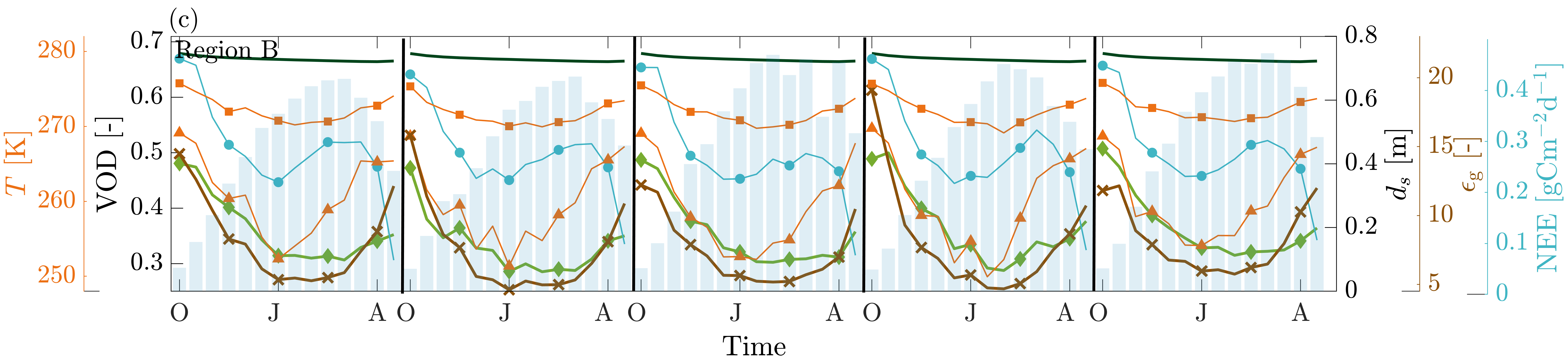}
    \caption{(a) Spatial extent of boreal forests over the continuous permafrost (Region A) as well as over isolated, sporadic, and discontinuous permafrost (Region B), and the time series of VOD, ground permittivity, NEE, ground ($T_{\rm g}$), and air temperature ($T_{\rm air}$) from October to April in five years 2015--2020. The dark green continuous line represents the climatology of VOD using NDVI data \citep{o2018algorithm}.}
 \label{fig:12}
\end{figure}

The temporal variability of the retrieved VOD and ground permittivity values are connected with other correlated variables such as ground temperature, 2-\si{m} air temperature, snow depth, and NEE. We focus our analyses on boreal forests over the continuous permafrost (Region A) as well as those over the discontinuous, sporadic, and isolated permafrost (Region B), shown by light and dark green shaded areas in Fig.~\ref{fig:12}\,a, respectively. Figs.~\ref{fig:12}\,b,c display the time series of spatial mean values of the variables of interest within each region from October to April 2015--2020, at a biweekly temporal resolution. 

As is evident, in both cases, the mean retrieved VOD, throughout the winter, is lower than the NDVI-derived climatological values -- especially over the discontinuous permafrost. We should admit that we have no direct ground validation data to reason about the existing shifts between the mean values. However, we need to recall that the NDVI-based VOD estimates are based on an empirical relationship \citep{jackson1999soil,jackson1991vegetation,o2018algorithm} in which VOD is a quadratic function of NDVI and hence has a positively skewed distribution, which can lead to an overestimation of NDVI-derived VOD over densely forested regions. Another important observation is that while the NDVI-derived VOD remains almost constant at around 0.3 (0.7) over continuous (discontinuous) permafrost, the retrieved VOD values show a significant temporal variation throughout the winter. This difference denotes that perhaps NDVI, as a surrogate variable, becomes less and less reliable for the estimation of VOD over higher latitudes.

Generally speaking, boreal forests over continuous permafrost consistently exhibit lower VOD values, declining from 0.4 in October to 0.2 in April, than over the discontinuous permafrost, where VOD drops from 0.5 in October to 0.3 in March and gradually begins to rise. This seems consistent with the gradual transition of the climate from a humid continental to a subarctic and tundra over higher latitudes and the fact that as the winter progresses, plants acclimatize to the colder air temperature by decreasing their transpiration, vegetation biomass, and xylem sap \citep{hincha2020introduction, schwank2021temperature}. At the same time, the differences observed after late March are attributed to the fact that colder air temperatures persist during a prolonged winter over continuous permafrost compared to discontinuous permafrost. For example, during the 2016-17 period, Region B had an increase in air temperature from 250 to 268~\si{K} and the ground temperature rose above the freezing point between early March and late April. However, in Region A, the air and ground temperatures remained predominantly below 270~\si{K} until late April.

The boreal forests over continuous permafrost exhibit lower values of ground permittivity (3.5--10) compared to those over discontinuous permafrost (5--15), which seems to be consistent with the climatology of the ground temperature over these two regions. Over the permanent permafrost, $\varepsilon_{\rm g}$ remains largely below 5 from late November to early April, indicating an almost frozen ground. However, Region B experiences $\varepsilon_{\rm g}>5$ for most of the time in winter except the month of January, signaling that the soil is not fully frozen for most of the time in winter despite the fact that on average a snow depth of 40~\si{cm} covers the ground.

\begin{wrapfigure}{r}{0.5\textwidth} \vspace{-10mm}
    \centering
    \includegraphics[width=1\linewidth]{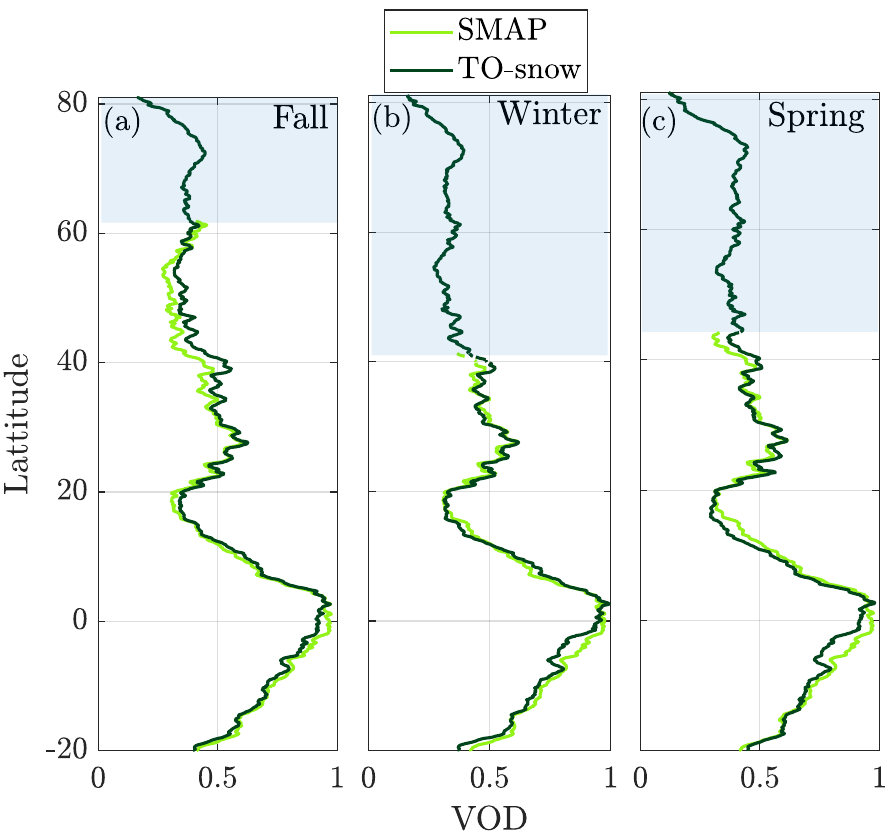}
    \caption{The latitudinal variations of zonal means of VOD retrieved using the TO-snow model shown against the official SMAP VOD product using the Dual Channel algorithm \citep{o2018algorithm}, for the seasons of Fall (Oct-Nov), Winter (Dec-Feb), and Spring (Mar-Apr) in the calendar year 2017-18. The blue color-shaded areas show the latitudes where more than 90\% of the pixels are covered with snow and hence retrievals only from the TO-snow model are reported.}
    \label{fig:13}
\end{wrapfigure}

It is interesting to note that the temporal dynamics of NEE and retrievals of VOD and $\varepsilon_{\rm g}$ are consistent in both regions. During the winter, as ground permittivity and VOD decrease, soil and plant respiration also decrease due to the reduced water content in soil and vegetation, leading to a decrease in NEE \citep{hunt1996global}, which is dominated by respiration rather than photosynthesis during the winter. For instance, over Region A, as ground permittivity and VOD decreases from 15 to 7 and 0.5 to 0.3, respectively, NEE also drops from 0.4 to 0.2~\si{gCm^{-2}d^{-1}}. Similarly, in Region B, as VOD decreases to 0.2 and ground permittivity to 3, NEE decreases to approximately 0.1~\si{gCm^{-2}d^{-1}}, which is expectedly lower than in Region B.  In both regions, NEE begins to increase with the rise in ground permittivity starting from early February as the soil warms up. However, in Region B, this increasing trend is followed by a sharp decrease in NEE in April, indicating the possibility of the earlier onset of photosynthesis highlighted by an increase in VOD compared to Region A where VOD does not begin to increase till late April. 

Furthermore, we compare the latitudinal variations of zonal means of VOD retrievals (Fig.~\ref{fig:13}) against those from SMAP products in the fall (Oct-Nov), winter (Dec-Feb), and spring (Mar-Apr) from 2017 to 2018. The results indicate reasonable agreement between the two retrievals over snow-free areas and show how the TO-snow model can extend the VOD retrievals to higher latitudes. The results seem to be consistent with the expected seasonal variations of the VOD as the minimum values over NH are reported during the winter and the difference seems to be more pronounced over the pan-Arctic lands above $60^\circ$~N.

\section{Conclusion}
\label{sec:V}

This study presented an improved version of the TO-snow model by \citet{kumawat2022passive} for the soil-snow-vegetation system at the L-band to reduce the complexity and computation time of the forward emission model. The emission model uses a two-layer composite radiative transfer framework to extend the classic tau-omega model for simulating the upwelling soil and reflected downwelling vegetation emissions in the presence of an intervening snow layer.  In this study, a comparative analysis is conducted between the outputs of the presented TO-snow model and the first-order 2-S emission model. A dual-channel inversion method was investigated to retrieve ground permittivity and VOD from SMAP radiometric observations, incorporating a priori information from reanalysis data about other unknown variables, such as snow density and 2-m air temperature.

Through controlled numerical experiments, it is demonstrated that in the presence of snow and moist soil, we are able to retrieve VOD (soil permittivity) with an error standard deviation not exceeding 0.1 (3.5) over a moderately dense vegetation canopy, with $\tau<$0.5, when the standard deviation of the soil moisture retrievals remains less than  0.04~\si{m^3.m^{-3}}. Using SMAP observations, the quality of both the TO-snow model and the inversion method was evaluated. Due to the lack of ground-based measurements for the retrieved variables over high-latitude snow-covered land surfaces, initial causal validation of the retrievals was conducted using vegetation proxies, NEE, and other relevant variables. While the presented causal validation results are encouraging, we acknowledge that further validation efforts are necessary. 

Given the objective of this research, we have employed an inversion method similar to the regularized DCA algorithm \citep{chaubell2021regularized}. It is emphasized that employing VOD climatology based on NDVI for retrievals over snow-covered regions is not the most suitable method due to the inherent uncertainty of NDVI measurements over high latitudes. Consequently, when dealing with snow-covered regions, it is presumed that the magnitude of the regularization parameter is notably lower than that of snow-free areas. Thus we reduced the parameter linearly to zero from 50\si{\degree}N poleward. 

It is worth noting that the presented TO-snow emission model holds potential for integration with different inversion techniques, including those that incorporate temporal constraints, such as the MT-DCA algorithm \citep{konings2016vegetation} and the SMAP-IB algorithm \citep{li2022new}. These inversion methods assume that VOD changes at a slower rate compared to soil moisture and hence ground permittivity, allowing for the assumption of near-constant VOD within a short time window. Conducting a comparative analysis to evaluate their performance is an intriguing area for future studies.

The presented emission model assumes a dry snowpack with no layering structure. However, to cover the entire range of land surface conditions and improve global modeling, future research needs to account for the impacts of a multilayered snowpack and especially snow wetness on the emission of the soil-snow system and evaluate the impacts on the retrieval uncertainties. The presence of liquid water in snow significantly increases its absorption, resulting in a decrease in L-band penetration depth from $>300$ m in dry snow to less than 3 (0.3)~m for snow with a liquid water content of ~1 (3)\% \citep{hofer1980investigations, matzler1984microwave}. 

In this paper, we made the assumption that the single scattering albedo ($\omega)$ remains the same as in the summer months. However, considering the decrease in vegetation water content during winter, it would be intriguing to investigate the impact of this decrease on $\omega$ in comparison to the summer months.

It is widely recognized that under frozen ground conditions, the L-band penetration depth increases in comparison to moist ground conditions, resulting in emissions originating from deeper layers of the soil \citep{lv2022simulation, lv2023impact}. In this study, we solely considered the effective temperature computed using the two-layer approach \citep{choudhury1982parameterization} where the weighted average of soil temperatures at the two soil layers, 5--15 and 15--35~\si{cm}, represents the effective temperature of the ground \citep{koster2020land}. These weights are parameterized corresponding to the experiments done for moist ground conditions \citep{wigneron2008estimating}. Hence, we implicitly overlooked any potential emissions from lower layers of frozen ground. Further investigation in this regard seems imperative. 

In this study, the focus was on retrieving VOD and ground permittivity for snow-covered areas. However, future research can be devoted to estimating the fraction of moist and frozen water content of partially frozen soils using relevant dielectric models such as the one by \citet{mironov2017temperature}. The percentage of frozen soil water depends on the total water content in the soil, bound water, clay fraction, ground temperature, and organic matter. Thus, given the ground permittivity, it is possible to retrieve freeze-thaw and the fraction of unfrozen water content in partially frozen soils. Expanding this idea to structurally and radiometrically complex permafrost soils within the SMAP footprint requires accounting for the sub-scale fractional abundance of (partially) frozen lakes, the spatial organization of the ice wedges, and the content of organic matter in soils.

\section{Data availability}
The codes and global dataset generated in this study, comprising VOD and ground permittivity measurements over snow-covered areas is openly accessible at the following location: \url{https://github.com/aebtehaj/SM-Snow-L-band}. 

\section{Author Contribution}
D. Kumawat and A. Ebtehaj conceptualized and designed the study. D. Kumawat conducted calculations, data analyses, and visualizations. All authors contributed to the discussion and edited the manuscript.

\section{Declaration of competing interest}
The authors declare that they have no conflict of interest.

\section{Acknowledgements}
The research is mainly supported by grants from NASA's Remote Sensing Theory program (RST, 80NSSC20K1717) through Dr. Lucia Tsaoussi and the Interdisciplinary Research in Earth Science program (IDS, 80NSSC20K1294) through Drs. Will McCarty and Aaron Pi\~{n}a.  Moreover, the authors acknowledge the Minnesota Supercomputing Institute (MSI, \url{http://www.msi.umn.edu}) at the University of Minnesota for providing computational resources that contributed to the research results reported in this paper. 

\clearpage
\appendix
\section{}
\label{A1}
\begin{equation}
    r^{\,p}_{\rm coh} = \left\lvert\frac{\xi_{\rm cs}^p+\tilde{\xi}_{\rm sg}^p\,e^{-2\gamma_{\rm s}d_{\rm s}\cos\alpha_{\rm s}}}{1+\xi_{\rm cs}^p\,\tilde{\xi}_{\rm sg}^p\,e^{-2\gamma_{\rm s}d_{\rm s}\cos\alpha_{\rm s}}}\right\rvert^2
\end{equation}

To calculate the mean value of the above expression, we simplify it algebraically by assuming zero attenuation constant for the snow layer, resulting in the following form:

\begin{equation}
    r^{\,p}_{\rm coh} = \frac{(\xi_{\rm cs}^p)^2 + (\tilde{\xi}_{\rm sg}^p)^2 +2\xi_{\rm cs}^p\tilde{\xi}_{\rm sg}^p\cos{\theta}}{1 + \xi_{\rm cs}^p\tilde{\xi}_{\rm sg}^p +2\xi_{\rm cs}^p\tilde{\xi}_{\rm sg}^p\cos{\theta}}, \quad \text{where} \quad \theta = -2\gamma_{\rm s}d_{\rm s}\cos\alpha_{\rm s}
\end{equation}

Integrating the above equation from $[0,2\pi]$ we get,
\begin{equation}
    r^{\,p}= \frac{1}{2\pi}\int_0^{2\pi} \frac{(\xi_{\rm cs}^p)^2 + (\tilde{\xi}_{\rm sg}^p)^2 +2\xi_{\rm cs}^p\tilde{\xi}_{\rm sg}^p\cos{\theta}}{1 + \xi_{\rm cs}^p\tilde{\xi}_{\rm sg}^p +2\xi_{\rm cs}^p\tilde{\xi}_{\rm sg}^p\cos{\theta}} d\theta
\end{equation}

Solving and simplifying the above equation, we get
\begin{equation}
     r^{\,p}  = 1+ \text{sgn}(|\xi_{\rm cs}^p| |\tilde{\xi}_{\rm sg}^p|-1)\left(\frac{|\xi_{\rm cs}^p|^2 + |\tilde{\xi}_{\rm sg}^p|^2-|\xi_{\rm cs}^p|^2 |\tilde{\xi}_{\rm sg}^p|^2-1}{|\xi_{\rm cs}^p|^2 |\tilde{\xi}_{\rm sg}^p|^2-1}\right) \\
\end{equation}

Similarly, we can obtain an expression for $e^p$ by simplifying the expression for $e^p_{\text{coh}}$. The simplified form of the expression is as follows:
\begin{equation}
    e^p = 1-r^p
\end{equation}


\clearpage
\bibliographystyle{plainnat}
\bibliography{refs.bib}

\begin{thebibliography}{95}
\providecommand{\natexlab}[1]{#1}
\providecommand{\url}[1]{\texttt{#1}}
\expandafter\ifx\csname urlstyle\endcsname\relax
  \providecommand{\doi}[1]{doi: #1}\else
  \providecommand{\doi}{doi: \begingroup \urlstyle{rm}\Url}\fi

\bibitem[Baldocchi(2008)]{baldocchi2008breathing}
Dennis Baldocchi.
\newblock ‘breathing’of the terrestrial biosphere: lessons learned from a
  global network of carbon dioxide flux measurement systems.
\newblock \emph{Australian Journal of Botany}, 56\penalty0 (1):\penalty0 1--26,
  2008.

\bibitem[Bircher et~al.(2016)Bircher, Demontoux, Razafindratsima, Zakharova,
  Drusch, Wigneron, and Kerr]{bircher2016band}
Simone Bircher, Fran{\c{c}}ois Demontoux, Stephen Razafindratsima, Elena
  Zakharova, Matthias Drusch, Jean-Pierre Wigneron, and Yann~H Kerr.
\newblock L-band relative permittivity of organic soil surface layers—a new
  dataset of resonant cavity measurements and model evaluation.
\newblock \emph{Remote Sensing}, 8\penalty0 (12):\penalty0 1024, 2016.

\bibitem[Bradshaw and Warkentin(2015)]{bradshaw2015global}
Corey~JA Bradshaw and Ian~G Warkentin.
\newblock Global estimates of boreal forest carbon stocks and flux.
\newblock \emph{Global and Planetary Change}, 128:\penalty0 24--30, 2015.

\bibitem[Brandt et~al.(2018)Brandt, Wigneron, Chave, Tagesson, Penuelas, Ciais,
  Rasmussen, Tian, Mbow, Al-Yaari, et~al.]{brandt2018satellite}
Martin Brandt, Jean-Pierre Wigneron, Jerome Chave, Torbern Tagesson, Josep
  Penuelas, Philippe Ciais, Kjeld Rasmussen, Feng Tian, Cheikh Mbow, Amen
  Al-Yaari, et~al.
\newblock Satellite passive microwaves reveal recent climate-induced carbon
  losses in african drylands.
\newblock \emph{Nature ecology \& evolution}, 2\penalty0 (5):\penalty0
  827--835, 2018.

\bibitem[Chan et~al.(2018)Chan, Bindlish, O'Neill, Jackson, Njoku, Dunbar,
  Chaubell, Piepmeier, Yueh, Entekhabi, Colliander, Chen, Cosh, Caldwell,
  Walker, Berg, McNairn, Thibeault, Mart{\'{i}}nez-Fern{\'{a}}ndez, Uldall,
  Seyfried, Bosch, Starks, {Holifield Collins}, Prueger, van~der Velde,
  Asanuma, Palecki, Small, Zreda, Calvet, Crow, and Kerr]{Chan2018}
S.~K. Chan, R.~Bindlish, P.~O'Neill, T.~Jackson, E.~Njoku, S.~Dunbar,
  J.~Chaubell, J.~Piepmeier, S.~Yueh, D.~Entekhabi, A.~Colliander, F.~Chen,
  M.~H. Cosh, T.~Caldwell, J.~Walker, A.~Berg, H.~McNairn, M.~Thibeault,
  J.~Mart{\'{i}}nez-Fern{\'{a}}ndez, F.~Uldall, M.~Seyfried, D.~Bosch,
  P.~Starks, C.~{Holifield Collins}, J.~Prueger, R.~van~der Velde, J.~Asanuma,
  M.~Palecki, E.~E. Small, M.~Zreda, J.~Calvet, W.~T. Crow, and Y.~Kerr.
\newblock {Development and assessment of the SMAP enhanced passive soil
  moisture product}.
\newblock \emph{Remote Sensing of Environment}, 204\penalty0 (October
  2017):\penalty0 931--941, 2018.
\newblock ISSN 00344257.
\newblock \doi{10.1016/j.rse.2017.08.025}.

\bibitem[Chan(2016)]{chan2016enhanced}
Steven Chan.
\newblock Enhanced level 3 passive soil moisture product specification
  document.
\newblock \emph{Jet Propulsion Laboratory}, 2016.

\bibitem[Chaparro et~al.(2022)Chaparro, Feldman, Chaubell, Yueh, and
  Entekhabi]{chaparro2022robustness}
David Chaparro, Andrew~F Feldman, Mario~Julian Chaubell, Simon~H Yueh, and Dara
  Entekhabi.
\newblock Robustness of vegetation optical depth retrievals based on l-band
  global radiometry.
\newblock \emph{IEEE Transactions on Geoscience and Remote Sensing},
  60:\penalty0 1--17, 2022.

\bibitem[Chaubell et~al.(2021)Chaubell, Yueh, Dunbar, Colliander, Entekhabi,
  Chan, Chen, Xu, Bindlish, O'Neill, et~al.]{chaubell2021regularized}
Julian Chaubell, Simon Yueh, R~Scott Dunbar, Andreas Colliander, Dara
  Entekhabi, Steven~K Chan, Fan Chen, Xiaolan Xu, Rajat Bindlish, Peggy
  O'Neill, et~al.
\newblock Regularized dual-channel algorithm for the retrieval of soil moisture
  and vegetation optical depth from smap measurements.
\newblock \emph{IEEE Journal of Selected Topics in Applied Earth Observations
  and Remote Sensing}, 15:\penalty0 102--114, 2021.

\bibitem[Choudhury et~al.(1979)Choudhury, Schmugge, Chang, and
  Newton]{choudhury1979effect}
BJ~Choudhury, Thomas~J Schmugge, A~Chang, and RW~Newton.
\newblock Effect of surface roughness on the microwave emission from soils.
\newblock \emph{Journal of Geophysical Research: Oceans}, 84\penalty0
  (C9):\penalty0 5699--5706, 1979.

\bibitem[Choudhury et~al.(1982)Choudhury, Schmugge, and
  Mo]{choudhury1982parameterization}
BJ~Choudhury, TJ~Schmugge, and T~Mo.
\newblock A parameterization of effective soil temperature for microwave
  emission.
\newblock \emph{Journal of Geophysical Research: Oceans}, 87\penalty0
  (C2):\penalty0 1301--1304, 1982.

\bibitem[Didan(2015)]{didan2015mod13c1}
K~Didan.
\newblock Mod13c1 modis/terra vegetation indices 16-day l3 global 0.05 deg cmg
  v006, lp daac--mod13c1 [data set], 2015.

\bibitem[Dou et~al.(2023)Dou, Tian, Wigneron, Tagesson, Du, Brandt, Liu, Zou,
  Kimball, and Fensholt]{dou2023reliability}
Yujie Dou, Feng Tian, Jean-Pierre Wigneron, Torbern Tagesson, Jinyang Du,
  Martin Brandt, Yi~Liu, Linqing Zou, John~S Kimball, and Rasmus Fensholt.
\newblock Reliability of using vegetation optical depth for estimating decadal
  and interannual carbon dynamics.
\newblock \emph{Remote Sensing of Environment}, 285:\penalty0 113390, 2023.

\bibitem[Eastman et~al.(2013)Eastman, Sangermano, Machado, Rogan, and
  Anyamba]{eastman2013global}
J~Ronald Eastman, Florencia Sangermano, Elia~A Machado, John Rogan, and Assaf
  Anyamba.
\newblock Global trends in seasonality of normalized difference vegetation
  index (ndvi), 1982--2011.
\newblock \emph{Remote Sensing}, 5\penalty0 (10):\penalty0 4799--4818, 2013.

\bibitem[Ebtehaj and Bras(2019)]{ebtehaj2019physically}
Ardeshir Ebtehaj and Rafael~L Bras.
\newblock A physically constrained inversion for high-resolution passive
  microwave retrieval of soil moisture and vegetation water content in l-band.
\newblock \emph{Remote Sensing of Environment}, 233:\penalty0 111346, 2019.

\bibitem[Entekhabi and Feldman(2019)]{entekhabi2019evaluating}
Dara Entekhabi and Andrew~F Feldman.
\newblock Evaluating brightness temperature information for estimating
  microwave land surface and vegetation properties.
\newblock In \emph{IGARSS 2019-2019 IEEE International Geoscience and Remote
  Sensing Symposium}, pages 5374--5377. IEEE, 2019.

\bibitem[Entekhabi et~al.(2010)Entekhabi, Njoku, O'Neill, Kellogg, Crow,
  Edelstein, Entin, Goodman, Jackson, Johnson, et~al.]{entekhabi2010soil}
Dara Entekhabi, Eni~G Njoku, Peggy~E O'Neill, Kent~H Kellogg, Wade~T Crow,
  Wendy~N Edelstein, Jared~K Entin, Shawn~D Goodman, Thomas~J Jackson, Joel
  Johnson, et~al.
\newblock The soil moisture active passive (smap) mission.
\newblock \emph{Proceedings of the IEEE}, 98\penalty0 (5):\penalty0 704--716,
  2010.

\bibitem[Frappart et~al.(2020)Frappart, Wigneron, Li, Liu, Al-Yaari, Fan, Wang,
  Moisy, Le~Masson, Aoulad~Lafkih, et~al.]{frappart2020global}
Fr{\'e}d{\'e}ric Frappart, Jean-Pierre Wigneron, Xiaojun Li, Xiangzhuo Liu,
  Amen Al-Yaari, Lei Fan, Mengjia Wang, Christophe Moisy, Erwan Le~Masson,
  Zacharie Aoulad~Lafkih, et~al.
\newblock Global monitoring of the vegetation dynamics from the vegetation
  optical depth (vod): A review.
\newblock \emph{Remote Sensing}, 12\penalty0 (18):\penalty0 2915, 2020.

\bibitem[Friedl et~al.(2002)Friedl, McIver, Hodges, Zhang, Muchoney, Strahler,
  Woodcock, Gopal, Schneider, Cooper, et~al.]{friedl2002global}
Mark~A Friedl, Douglas~K McIver, John~CF Hodges, Xiaoyang~Y Zhang, D~Muchoney,
  Alan~H Strahler, Curtis~E Woodcock, Sucharita Gopal, Annemarie Schneider,
  Amanda Cooper, et~al.
\newblock Global land cover mapping from modis: algorithms and early results.
\newblock \emph{Remote sensing of Environment}, 83\penalty0 (1-2):\penalty0
  287--302, 2002.

\bibitem[Gao et~al.(2020{\natexlab{a}})Gao, Sadeghi, and
  Ebtehaj]{gao2020microwave}
Lun Gao, Morteza Sadeghi, and Ardeshir Ebtehaj.
\newblock Microwave retrievals of soil moisture and vegetation optical depth
  with improved resolution using a combined constrained inversion algorithm:
  Application for smap satellite.
\newblock \emph{Remote Sensing of Environment}, 239:\penalty0 111662,
  2020{\natexlab{a}}.

\bibitem[Gao et~al.(2020{\natexlab{b}})Gao, Sadeghi, Feldman, and
  Ebtehaj]{gao2020spatially}
Lun Gao, Morteza Sadeghi, Andrew~F Feldman, and Ardeshir Ebtehaj.
\newblock A spatially constrained multichannel algorithm for inversion of a
  first-order microwave emission model at l-band.
\newblock \emph{IEEE Transactions on Geoscience and Remote Sensing},
  58\penalty0 (11):\penalty0 8134--8146, 2020{\natexlab{b}}.

\bibitem[Gao et~al.(2021)Gao, Ebtehaj, Chaubell, Sadeghi, Li, and
  Wigneron]{gao2021reappraisal}
Lun Gao, Ardeshir Ebtehaj, Mario~Julian Chaubell, Morteza Sadeghi, Xiaojun Li,
  and Jean-Pierre Wigneron.
\newblock Reappraisal of smap inversion algorithms for soil moisture and
  vegetation optical depth.
\newblock \emph{Remote Sensing of Environment}, 264:\penalty0 112627, 2021.

\bibitem[Gao et~al.(2022)Gao, Ebtehaj, Cohen, and Wigneron]{gao2022variability}
Lun Gao, Ardeshir Ebtehaj, Judah Cohen, and Jean-Pierre Wigneron.
\newblock Variability and changes of unfrozen soils below snowpack.
\newblock \emph{Geophysical Research Letters}, 49\penalty0 (4):\penalty0
  e2021GL095354, 2022.

\bibitem[Hallikainen et~al.(1986)Hallikainen, Ulaby, and
  Abdelrazik]{hallikainen1986dielectric}
Martti Hallikainen, F~Ulaby, and Mohamed Abdelrazik.
\newblock Dielectric properties of snow in the 3 to 37 ghz range.
\newblock \emph{IEEE transactions on Antennas and Propagation}, 34\penalty0
  (11):\penalty0 1329--1340, 1986.

\bibitem[Hersbach et~al.(2018)Hersbach, Bell, Berrisford, Biavati, Horányi,
  Mu{\~n}oz~Sabater, et~al.]{hersbach2018copernicus}
H~Hersbach, B~Bell, P~Berrisford, G~Biavati, A~Horányi, J~Mu{\~n}oz~Sabater,
  et~al.
\newblock Copernicus climate change service (c3s) climate data store (cds).
\newblock \emph{ERA5 hourly data on single levels from 1979 to present}, 2018.

\bibitem[Hincha and Zuther(2020)]{hincha2020introduction}
Dirk~K Hincha and Ellen Zuther.
\newblock Introduction: plant cold acclimation and winter survival.
\newblock \emph{Plant Cold Acclimation: Methods and Protocols}, pages 1--7,
  2020.

\bibitem[Hofer and M{\"a}tzler(1980)]{hofer1980investigations}
R~Hofer and Ch~M{\"a}tzler.
\newblock Investigations on snow parameters by radiometry in the 3-to 60-mm
  wavelength region.
\newblock \emph{Journal of Geophysical Research: Oceans}, 85\penalty0
  (C1):\penalty0 453--460, 1980.

\bibitem[Houghton(2005)]{houghton2005aboveground}
RA~Houghton.
\newblock Aboveground forest biomass and the global carbon balance.
\newblock \emph{Global change biology}, 11\penalty0 (6):\penalty0 945--958,
  2005.

\bibitem[Hugelius et~al.(2014)Hugelius, Strauss, Zubrzycki, Harden, Schuur,
  Ping, Schirrmeister, Grosse, Michaelson, Koven,
  et~al.]{hugelius2014estimated}
Gustaf Hugelius, Jens Strauss, Sebastian Zubrzycki, Jennifer~W Harden,
  Edward~AG Schuur, C-L Ping, Lutz Schirrmeister, Guido Grosse, Gary~J
  Michaelson, Charles~D Koven, et~al.
\newblock Estimated stocks of circumpolar permafrost carbon with quantified
  uncertainty ranges and identified data gaps.
\newblock \emph{Biogeosciences}, 11\penalty0 (23):\penalty0 6573--6593, 2014.

\bibitem[Hunt~Jr et~al.(1996)Hunt~Jr, Piper, Nemani, Keeling, Otto, and
  Running]{hunt1996global}
E~Raymond Hunt~Jr, Stephen~C Piper, Ramakrishna Nemani, Charles~D Keeling,
  Ralf~D Otto, and Steven~W Running.
\newblock Global net carbon exchange and intra-annual atmospheric co2
  concentrations predicted by an ecosystem process model and three-dimensional
  atmospheric transport model.
\newblock \emph{Global Biogeochemical Cycles}, 10\penalty0 (3):\penalty0
  431--456, 1996.

\bibitem[Jackson et~al.(1999)Jackson, Le~Vine, Hsu, Oldak, Starks, Swift,
  Isham, and Haken]{jackson1999soil}
Thomas~J Jackson, David~M Le~Vine, Ann~Y Hsu, Anna Oldak, Patrick~J Starks,
  Calvin~T Swift, John~D Isham, and Michael Haken.
\newblock Soil moisture mapping at regional scales using microwave radiometry:
  The southern great plains hydrology experiment.
\newblock \emph{IEEE transactions on geoscience and remote sensing},
  37\penalty0 (5):\penalty0 2136--2151, 1999.

\bibitem[Jackson and Schmugge(1991)]{jackson1991vegetation}
TJ~Jackson and TJ~Schmugge.
\newblock Vegetation effects on the microwave emission of soils.
\newblock \emph{Remote Sensing of Environment}, 36\penalty0 (3):\penalty0
  203--212, 1991.

\bibitem[Jia et~al.(2009)Jia, Epstein, and Walker]{jia2009vegetation}
Gensuo~J Jia, Howard~E Epstein, and Donald~A Walker.
\newblock Vegetation greening in the canadian arctic related to decadal
  warming.
\newblock \emph{Journal of Environmental Monitoring}, 11\penalty0
  (12):\penalty0 2231--2238, 2009.

\bibitem[Jin et~al.(2021)Jin, Jin, Iwahana, Marchenko, Luo, Li, and
  Liang]{jin2021impacts}
Xiao-Ying Jin, Hui-Jun Jin, Go~Iwahana, Sergey~S Marchenko, Dong-Liang Luo,
  Xiao-Ying Li, and Si-Hai Liang.
\newblock Impacts of climate-induced permafrost degradation on vegetation: A
  review.
\newblock \emph{Advances in Climate Change Research}, 12\penalty0 (1):\penalty0
  29--47, 2021.

\bibitem[Jung et~al.(2009)Jung, Reichstein, and Bondeau]{jung2009towards}
M~Jung, M~Reichstein, and Alberte Bondeau.
\newblock Towards global empirical upscaling of fluxnet eddy covariance
  observations: validation of a model tree ensemble approach using a biosphere
  model.
\newblock \emph{Biogeosciences}, 6\penalty0 (10):\penalty0 2001--2013, 2009.

\bibitem[Jung et~al.(2011)Jung, Reichstein, Margolis, Cescatti, Richardson,
  Arain, Arneth, Bernhofer, Bonal, Chen, et~al.]{jung2011global}
Martin Jung, Markus Reichstein, Hank~A Margolis, Alessandro Cescatti, Andrew~D
  Richardson, M~Altaf Arain, Almut Arneth, Christian Bernhofer, Damien Bonal,
  Jiquan Chen, et~al.
\newblock Global patterns of land-atmosphere fluxes of carbon dioxide, latent
  heat, and sensible heat derived from eddy covariance, satellite, and
  meteorological observations.
\newblock \emph{Journal of Geophysical Research: Biogeosciences}, 116\penalty0
  (G3), 2011.

\bibitem[Jung et~al.(2019)Jung, Koirala, Weber, Ichii, Gans, Camps-Valls,
  Papale, Schwalm, Tramontana, and Reichstein]{jung2019fluxcom}
Martin Jung, Sujan Koirala, Ulrich Weber, Kazuhito Ichii, Fabian Gans, Gustau
  Camps-Valls, Dario Papale, Christopher Schwalm, Gianluca Tramontana, and
  Markus Reichstein.
\newblock The fluxcom ensemble of global land-atmosphere energy fluxes.
\newblock \emph{Scientific data}, 6\penalty0 (1):\penalty0 74, 2019.

\bibitem[Kerr et~al.(2010)Kerr, Waldteufel, Wigneron, Delwart, Cabot, Boutin,
  Escorihuela, Font, Reul, Gruhier, et~al.]{kerr2010smos}
Yann~H Kerr, Philippe Waldteufel, Jean-Pierre Wigneron, Steven Delwart,
  Fran{\c{c}}ois Cabot, Jacqueline Boutin, Maria-Jos{\'e} Escorihuela, Jordi
  Font, Nicolas Reul, Claire Gruhier, et~al.
\newblock The smos mission: New tool for monitoring key elements ofthe global
  water cycle.
\newblock \emph{Proceedings of the IEEE}, 98\penalty0 (5):\penalty0 666--687,
  2010.

\bibitem[Konings et~al.(2016)Konings, Piles, R{\"o}tzer, McColl, Chan, and
  Entekhabi]{konings2016vegetation}
Alexandra~G Konings, Mar{\'\i}a Piles, Kathrina R{\"o}tzer, Kaighin~A McColl,
  Steven~K Chan, and Dara Entekhabi.
\newblock Vegetation optical depth and scattering albedo retrieval using time
  series of dual-polarized l-band radiometer observations.
\newblock \emph{Remote sensing of environment}, 172:\penalty0 178--189, 2016.

\bibitem[Koster et~al.(2020)Koster, Reichle, Mahanama, Perket, Liu, and
  Partyka]{koster2020land}
Randal~D Koster, Rolf~H Reichle, Sarith~PP Mahanama, Justin Perket, Qing Liu,
  and Gary Partyka.
\newblock Land-focused changes in the updated geos fp system (version 5.25).
\newblock Technical report, 2020.

\bibitem[Kumawat et~al.(2022)Kumawat, Olyaei, Gao, and
  Ebtehaj]{kumawat2022passive}
Divya Kumawat, Mohammadali Olyaei, Lun Gao, and Ardeshir Ebtehaj.
\newblock Passive microwave retrieval of soil moisture below snowpack at l-band
  using smap observations.
\newblock \emph{IEEE Transactions on Geoscience and Remote Sensing},
  60:\penalty0 1--16, 2022.

\bibitem[Lemmetyinen et~al.(2016)Lemmetyinen, Schwank, Rautiainen, Kontu,
  Parkkinen, M{\"a}tzler, Wiesmann, Wegm{\"u}ller, Derksen, Toose,
  et~al.]{lemmetyinen2016snow}
Juha Lemmetyinen, Mike Schwank, Kimmo Rautiainen, Anna Kontu, Tiina Parkkinen,
  Christian M{\"a}tzler, Andreas Wiesmann, Urs Wegm{\"u}ller, Chris Derksen,
  Peter Toose, et~al.
\newblock Snow density and ground permittivity retrieved from l-band
  radiometry: Application to experimental data.
\newblock \emph{Remote sensing of environment}, 180:\penalty0 377--391, 2016.

\bibitem[Li et~al.(2021)Li, Wigneron, Frappart, Fan, Ciais, Fensholt,
  Entekhabi, Brandt, Konings, Liu, et~al.]{li2021global}
Xiaojun Li, Jean-Pierre Wigneron, Fr{\'e}d{\'e}ric Frappart, Lei Fan, Philippe
  Ciais, Rasmus Fensholt, Dara Entekhabi, Martin Brandt, Alexandra~G Konings,
  Xiangzhuo Liu, et~al.
\newblock Global-scale assessment and inter-comparison of recently
  developed/reprocessed microwave satellite vegetation optical depth products.
\newblock \emph{Remote Sensing of Environment}, 253:\penalty0 112208, 2021.

\bibitem[Li et~al.(2022)Li, Wigneron, Fan, Frappart, Yueh, Colliander, Ebtehaj,
  Gao, Fernandez-Moran, Liu, et~al.]{li2022new}
Xiaojun Li, Jean-Pierre Wigneron, Lei Fan, Fr{\'e}d{\'e}ric Frappart, Simon~H
  Yueh, Andreas Colliander, Ardeshir Ebtehaj, Lun Gao, Roberto Fernandez-Moran,
  Xiangzhuo Liu, et~al.
\newblock A new smap soil moisture and vegetation optical depth product
  (smap-ib): Algorithm, assessment and inter-comparison.
\newblock \emph{Remote Sensing of Environment}, 271:\penalty0 112921, 2022.

\bibitem[Liu et~al.(2015)Liu, Van~Dijk, De~Jeu, Canadell, McCabe, Evans, and
  Wang]{liu2015recent}
Yi~Y Liu, Albert~IJM Van~Dijk, Richard~AM De~Jeu, Josep~G Canadell, Matthew~F
  McCabe, Jason~P Evans, and Guojie Wang.
\newblock Recent reversal in loss of global terrestrial biomass.
\newblock \emph{Nature Climate Change}, 5\penalty0 (5):\penalty0 470--474,
  2015.

\bibitem[Loveland et~al.(2000)Loveland, Reed, Brown, Ohlen, Zhu, Yang, and
  Merchant]{loveland2000development}
Thomas~R Loveland, Bradley~C Reed, Jesslyn~F Brown, Donald~O Ohlen, Zhiliang
  Zhu, LWMJ Yang, and James~W Merchant.
\newblock Development of a global land cover characteristics database and igbp
  discover from 1 km avhrr data.
\newblock \emph{International journal of remote sensing}, 21\penalty0
  (6-7):\penalty0 1303--1330, 2000.

\bibitem[L{\"u}ers et~al.(2014)L{\"u}ers, Westermann, Piel, and
  Boike]{luers2014annual}
Johannes L{\"u}ers, Sebastian Westermann, Konstanze Piel, and Julia Boike.
\newblock Annual co 2 budget and seasonal co 2 exchange signals at a high
  arctic permafrost site on spitsbergen, svalbard archipelago.
\newblock \emph{Biogeosciences}, 11\penalty0 (22):\penalty0 6307--6322, 2014.

\bibitem[Lv et~al.(2022)Lv, Simmer, Zeng, Wen, and Su]{lv2022simulation}
Shaoning Lv, Clemens Simmer, Yijian Zeng, Jun Wen, and Zhongbo Su.
\newblock The simulation of l-band microwave emission of frozen soil during the
  thawing period with the community microwave emission model (cmem).
\newblock \emph{Journal of Remote Sensing}, 2022, 2022.

\bibitem[Lv et~al.(2023)Lv, Simmer, Zeng, Su, and Wen]{lv2023impact}
Shaoning Lv, Clemens Simmer, Yijian Zeng, Zhongbo Su, and Jun Wen.
\newblock Impact of profile-averaged soil ice fraction on passive microwave
  brightness temperature diurnal amplitude variations (dav) at l-band.
\newblock \emph{Cold Regions Science and Technology}, 205:\penalty0 103674,
  2023.

\bibitem[Matzler et~al.(1984)Matzler, Aebischer, and
  Schanda]{matzler1984microwave}
C~Matzler, H~Aebischer, and E~Schanda.
\newblock Microwave dielectric properties of surface snow.
\newblock \emph{IEEE Journal of Oceanic Engineering}, 9\penalty0 (5):\penalty0
  366--371, 1984.

\bibitem[M{\"a}tzler(1998)]{matzler1998improved}
Christian M{\"a}tzler.
\newblock Improved born approximation for scattering of radiation in a granular
  medium.
\newblock \emph{Journal of Applied Physics}, 83\penalty0 (11):\penalty0
  6111--6117, 1998.

\bibitem[M{\"a}tzler(2006)]{matzler2006thermal}
Christian M{\"a}tzler.
\newblock \emph{Thermal microwave radiation: applications for remote sensing},
  volume~52.
\newblock Iet, 2006.

\bibitem[Mavrovic et~al.(2021)Mavrovic, Pardo~Lara, Berg, Demontoux, Royer, and
  Roy]{mavrovic2021soil}
Alex Mavrovic, Renato Pardo~Lara, Aaron Berg, Fran{\c{c}}ois Demontoux, Alain
  Royer, and Alexandre Roy.
\newblock Soil dielectric characterization during freeze--thaw transitions
  using l-band coaxial and soil moisture probes.
\newblock \emph{Hydrology and Earth System Sciences}, 25\penalty0 (3):\penalty0
  1117--1131, 2021.

\bibitem[McGuire et~al.(2018)McGuire, Lawrence, Koven, Clein, Burke, Chen,
  Jafarov, MacDougall, Marchenko, Nicolsky, et~al.]{mcguire2018dependence}
A~David McGuire, David~M Lawrence, Charles Koven, Joy~S Clein, Eleanor Burke,
  Guangsheng Chen, Elchin Jafarov, Andrew~H MacDougall, Sergey Marchenko,
  Dmitry Nicolsky, et~al.
\newblock Dependence of the evolution of carbon dynamics in the northern
  permafrost region on the trajectory of climate change.
\newblock \emph{Proceedings of the National Academy of Sciences}, 115\penalty0
  (15):\penalty0 3882--3887, 2018.

\bibitem[Mironov et~al.(2009)Mironov, Kosolapova, and
  Fomin]{mironov2009physically}
Valery~L Mironov, Lyudmila~G Kosolapova, and Sergej~V Fomin.
\newblock Physically and mineralogically based spectroscopic dielectric model
  for moist soils.
\newblock \emph{IEEE Transactions on Geoscience and Remote Sensing},
  47\penalty0 (7):\penalty0 2059--2070, 2009.

\bibitem[Mironov et~al.(2017)Mironov, Kosolapova, Lukin, Karavaysky, and
  Molostov]{mironov2017temperature}
Valery~L Mironov, Liudmila~G Kosolapova, Yury~I Lukin, Andrey~Y Karavaysky, and
  Illia~P Molostov.
\newblock Temperature-and texture-dependent dielectric model for frozen and
  thawed mineral soils at a frequency of 1.4 ghz.
\newblock \emph{Remote Sensing of Environment}, 200:\penalty0 240--249, 2017.

\bibitem[Mo et~al.(1982)Mo, Choudhury, Schmugge, Wang, and
  Jackson]{mo1982model}
T~Mo, BJ~Choudhury, TJ~Schmugge, James~R Wang, and TJ~Jackson.
\newblock A model for microwave emission from vegetation-covered fields.
\newblock \emph{Journal of Geophysical Research: Oceans}, 87\penalty0
  (C13):\penalty0 11229--11237, 1982.

\bibitem[Mu{\~n}oz-Sabater et~al.(2021)Mu{\~n}oz-Sabater, Dutra,
  Agust{\'\i}-Panareda, Albergel, Arduini, Balsamo, Boussetta, Choulga,
  Harrigan, Hersbach, et~al.]{munoz2021era5}
Joaqu{\'\i}n Mu{\~n}oz-Sabater, Emanuel Dutra, Anna Agust{\'\i}-Panareda,
  Cl{\'e}ment Albergel, Gabriele Arduini, Gianpaolo Balsamo, Souhail Boussetta,
  Margarita Choulga, Shaun Harrigan, Hans Hersbach, et~al.
\newblock Era5-land: A state-of-the-art global reanalysis dataset for land
  applications.
\newblock \emph{Earth System Science Data}, 13\penalty0 (9):\penalty0
  4349--4383, 2021.

\bibitem[Naderpour et~al.(2017{\natexlab{a}})Naderpour, Schwank, and
  M{\"a}tzler]{naderpour2017davos}
Reza Naderpour, Mike Schwank, and Christian M{\"a}tzler.
\newblock Davos-laret remote sensing field laboratory: 2016/2017 winter season
  l-band measurements data-processing and analysis.
\newblock \emph{Remote Sensing}, 9\penalty0 (11):\penalty0 1185,
  2017{\natexlab{a}}.

\bibitem[Naderpour et~al.(2017{\natexlab{b}})Naderpour, Schwank, M{\"a}tzler,
  Lemmetyinen, and Steffen]{naderpour2017snow}
Reza Naderpour, Mike Schwank, Christian M{\"a}tzler, Juha Lemmetyinen, and
  Konrad Steffen.
\newblock Snow density and ground permittivity retrieved from l-band
  radiometry: A retrieval sensitivity analysis.
\newblock \emph{IEEE Journal of selected topics in applied earth observations
  and remote sensing}, 10\penalty0 (7):\penalty0 3148--3161,
  2017{\natexlab{b}}.

\bibitem[Naderpour et~al.(2022)Naderpour, Schwank, Houtz, and
  M{\"a}tzler]{naderpour2022band}
Reza Naderpour, Mike Schwank, Derek Houtz, and Christian M{\"a}tzler.
\newblock L-band radiometry of alpine seasonal snow cover: 4 years at the
  davos-laret remote sensing field laboratory.
\newblock \emph{IEEE Journal of Selected Topics in Applied Earth Observations
  and Remote Sensing}, 15:\penalty0 8199--8220, 2022.

\bibitem[{Obu} et~al.(2018){Obu}, {Westermann}, {K\"{a}\"{a}b}, and
  {Bartsch}]{obu2018gtmn}
Jaroslav {Obu}, Sebastian {Westermann}, Andreas {K\"{a}\"{a}b}, and Annett
  {Bartsch}.
\newblock {Ground Temperature Map, 2000-2016, Northern Hemisphere Permafrost},
  2018.
\newblock URL \url{https://doi.org/10.1594/PANGAEA.888600}.

\bibitem[O'Neill et~al.(2020)O'Neill, Bindlish, Chan, Njoku, and
  Jackson]{o2018algorithm}
Peggy O'Neill, Rajat Bindlish, Steven Chan, Eni Njoku, and Tom Jackson.
\newblock Algorithm theoretical basis document. level 2 \& 3 soil moisture
  (passive) data products.
\newblock 2020.

\bibitem[Piao et~al.(2020)Piao, Wang, Park, Chen, Lian, He, Bjerke, Chen,
  Ciais, T{\o}mmervik, et~al.]{piao2020characteristics}
Shilong Piao, Xuhui Wang, Taejin Park, Chi Chen, XU~Lian, Yue He, Jarle~W
  Bjerke, Anping Chen, Philippe Ciais, Hans T{\o}mmervik, et~al.
\newblock Characteristics, drivers and feedbacks of global greening.
\newblock \emph{Nature Reviews Earth \& Environment}, 1\penalty0 (1):\penalty0
  14--27, 2020.

\bibitem[Potapov et~al.(2022)Potapov, Hansen, Pickens, Hernandez-Serna,
  Tyukavina, Turubanova, Zalles, Li, Khan, Stolle, et~al.]{potapov2022global}
P~Potapov, MC~Hansen, A~Pickens, A~Hernandez-Serna, A~Tyukavina, S~Turubanova,
  V~Zalles, X~Li, A~Khan, F~Stolle, et~al.
\newblock The global 2000--2020 land cover and land use change dataset derived
  from the landsat archive: first results.
\newblock \emph{Front. Remote Sens}, 3, 2022.

\bibitem[Potapov et~al.(2008)Potapov, Hansen, Stehman, Loveland, and
  Pittman]{potapov2008combining}
Peter Potapov, Matthew~C Hansen, Stephen~V Stehman, Thomas~R Loveland, and Kyle
  Pittman.
\newblock Combining modis and landsat imagery to estimate and map boreal forest
  cover loss.
\newblock \emph{Remote sensing of environment}, 112\penalty0 (9):\penalty0
  3708--3719, 2008.

\bibitem[Rees et~al.(2020)Rees, Hofgaard, Boudreau, Cairns, Harper, Mamet,
  Mathisen, Swirad, and Tutubalina]{rees2020subarctic}
W~Gareth Rees, Annika Hofgaard, St{\'e}phane Boudreau, David~M Cairns, Karen
  Harper, Steven Mamet, Ingrid Mathisen, Zuzanna Swirad, and Olga Tutubalina.
\newblock Is subarctic forest advance able to keep pace with climate change?
\newblock \emph{Global Change Biology}, 26\penalty0 (7):\penalty0 3965--3977,
  2020.

\bibitem[Richardson et~al.(2013)Richardson, Keenan, Migliavacca, Ryu,
  Sonnentag, and Toomey]{richardson2013climate}
Andrew~D Richardson, Trevor~F Keenan, Mirco Migliavacca, Youngryel Ryu, Oliver
  Sonnentag, and Michael Toomey.
\newblock Climate change, phenology, and phenological control of vegetation
  feedbacks to the climate system.
\newblock \emph{Agricultural and Forest Meteorology}, 169:\penalty0 156--173,
  2013.

\bibitem[Rodr{\'\i}guez-Fern{\'a}ndez et~al.(2018)Rodr{\'\i}guez-Fern{\'a}ndez,
  Mialon, Mermoz, Bouvet, Richaume, Al~Bitar, Al-Yaari, Brandt, Kaminski,
  Le~Toan, et~al.]{rodriguez2018evaluation}
Nemesio~J Rodr{\'\i}guez-Fern{\'a}ndez, Arnaud Mialon, Stephane Mermoz,
  Alexandre Bouvet, Philippe Richaume, Ahmad Al~Bitar, Amen Al-Yaari, Martin
  Brandt, Thomas Kaminski, Thuy Le~Toan, et~al.
\newblock An evaluation of smos l-band vegetation optical depth (l-vod) data
  sets: high sensitivity of l-vod to above-ground biomass in africa.
\newblock \emph{Biogeosciences}, 15\penalty0 (14):\penalty0 4627--4645, 2018.

\bibitem[Romanovsky et~al.(2010)Romanovsky, Drozdov, Oberman, Malkova,
  Kholodov, Marchenko, Moskalenko, Sergeev, Ukraintseva, Abramov,
  et~al.]{romanovsky2010thermal}
Vladimir~E Romanovsky, DS~Drozdov, Naum~G Oberman, GV~Malkova, Alexander~L
  Kholodov, SS~Marchenko, Nataliya~G Moskalenko, DO~Sergeev, NG~Ukraintseva,
  AA~Abramov, et~al.
\newblock Thermal state of permafrost in russia.
\newblock \emph{Permafrost and Periglacial Processes}, 21\penalty0
  (2):\penalty0 136--155, 2010.

\bibitem[Schuur et~al.(2022)Schuur, Abbott, Commane, Ernakovich, Euskirchen,
  Hugelius, Grosse, Jones, Koven, Leshyk, et~al.]{schuur2022permafrost}
Edward~AG Schuur, Benjamin~W Abbott, Roisin Commane, Jessica Ernakovich,
  Eugenie Euskirchen, Gustaf Hugelius, Guido Grosse, Miriam Jones, Charlie
  Koven, Victor Leshyk, et~al.
\newblock Permafrost and climate change: carbon cycle feedbacks from the
  warming arctic.
\newblock \emph{Annual Review of Environment and Resources}, 47:\penalty0
  343--371, 2022.

\bibitem[Schwank et~al.(2014)Schwank, Rautiainen, M{\"a}tzler, St{\"a}hli,
  Lemmetyinen, Pulliainen, Vehvil{\"a}inen, Kontu, Ikonen, M{\'e}nard,
  et~al.]{schwank2014model}
Mike Schwank, Kimmo Rautiainen, Christian M{\"a}tzler, Manfred St{\"a}hli, Juha
  Lemmetyinen, Jouni Pulliainen, Juho Vehvil{\"a}inen, Anna Kontu, Jaakko
  Ikonen, C{\'e}cile~B M{\'e}nard, et~al.
\newblock Model for microwave emission of a snow-covered ground with focus on l
  band.
\newblock \emph{Remote sensing of environment}, 154:\penalty0 180--191, 2014.

\bibitem[Schwank et~al.(2015)Schwank, M{\"a}tzler, Wiesmann, Wegm{\"u}ller,
  Pulliainen, Lemmetyinen, Rautiainen, Derksen, Toose, and
  Drusch]{schwank2015snow}
Mike Schwank, Christian M{\"a}tzler, Andreas Wiesmann, Urs Wegm{\"u}ller, Jouni
  Pulliainen, Juha Lemmetyinen, Kimmo Rautiainen, Chris Derksen, Peter Toose,
  and Matthias Drusch.
\newblock Snow density and ground permittivity retrieved from l-band
  radiometry: A synthetic analysis.
\newblock \emph{IEEE Journal of Selected Topics in Applied Earth Observations
  and Remote Sensing}, 8\penalty0 (8):\penalty0 3833--3845, 2015.

\bibitem[Schwank et~al.(2018)Schwank, Naderpour, and
  M{\"a}tzler]{schwank2018tau}
Mike Schwank, Reza Naderpour, and Christian M{\"a}tzler.
\newblock “tau-omega”-and two-stream emission models used for passive
  l-band retrievals: Application to close-range measurements over a forest.
\newblock \emph{Remote Sensing}, 10\penalty0 (12):\penalty0 1868, 2018.

\bibitem[Schwank et~al.(2021)Schwank, Kontu, Mialon, Naderpour, Houtz,
  Lemmetyinen, Rautiainen, Li, Richaume, Kerr, et~al.]{schwank2021temperature}
Mike Schwank, Anna Kontu, Arnaud Mialon, Reza Naderpour, Derek Houtz, Juha
  Lemmetyinen, Kimmo Rautiainen, Qinghuan Li, Philippe Richaume, Yann Kerr,
  et~al.
\newblock Temperature effects on l-band vegetation optical depth of a boreal
  forest.
\newblock \emph{Remote Sensing of Environment}, 263:\penalty0 112542, 2021.

\bibitem[Shorohova et~al.(2009)Shorohova, Kuuluvainen, Kangur, and
  J{\~o}giste]{shorohova2009natural}
Ekaterina Shorohova, Timo Kuuluvainen, Ahto Kangur, and Kalev J{\~o}giste.
\newblock Natural stand structures, disturbance regimes and successional
  dynamics in the eurasian boreal forests: a review with special reference to
  russian studies.
\newblock \emph{Annals of Forest Science}, 66\penalty0 (2):\penalty0 1--20,
  2009.

\bibitem[Simard et~al.(2011)Simard, Pinto, Fisher, and
  Baccini]{simard2011mapping}
Marc Simard, Naiara Pinto, Joshua~B Fisher, and Alessandro Baccini.
\newblock Mapping forest canopy height globally with spaceborne lidar.
\newblock \emph{Journal of Geophysical Research: Biogeosciences}, 116\penalty0
  (G4), 2011.

\bibitem[Sturm and Liston(2021)]{sturm2021revisiting}
Matthew Sturm and Glen~E Liston.
\newblock Revisiting the global seasonal snow classification: An updated
  dataset for earth system applications.
\newblock \emph{Journal of Hydrometeorology}, 22\penalty0 (11):\penalty0
  2917--2938, 2021.

\bibitem[Sutinen et~al.(2008)Sutinen, H{\"a}nninen, and
  Ven{\"a}l{\"a}inen]{sutinen2008effect}
Raimo Sutinen, Pekka H{\"a}nninen, and Ari Ven{\"a}l{\"a}inen.
\newblock Effect of mild winter events on soil water content beneath snowpack.
\newblock \emph{Cold Regions Science and Technology}, 51\penalty0 (1):\penalty0
  56--67, 2008.

\bibitem[Tamang et~al.(2019)Tamang, Ebtehaj, Prein, and Heymsfield]{Tamang2019}
Sagar~K. Tamang, Ardeshir~M. Ebtehaj, Andreas~F. Prein, and Andrew~J.
  Heymsfield.
\newblock Linking global changes of snowfall and wet-bulb temperature.
\newblock \emph{Journal of Climate}, 9 2019.
\newblock ISSN 0894-8755.
\newblock \doi{10.1175/jcli-d-19-0254.1}.

\bibitem[Teubner et~al.(2018)Teubner, Forkel, Jung, Liu, Miralles, Parinussa,
  Van~der Schalie, Vreugdenhil, Schwalm, Tramontana,
  et~al.]{teubner2018assessing}
Irene~E Teubner, Matthias Forkel, Martin Jung, Yi~Y Liu, Diego~G Miralles,
  Robert Parinussa, Robin Van~der Schalie, Mariette Vreugdenhil, Christopher~R
  Schwalm, Gianluca Tramontana, et~al.
\newblock Assessing the relationship between microwave vegetation optical depth
  and gross primary production.
\newblock \emph{International journal of applied earth observation and
  geoinformation}, 65:\penalty0 79--91, 2018.

\bibitem[Tian et~al.(2018)Tian, Wigneron, Ciais, Chave, Og{\'e}e, Pe{\~n}uelas,
  R{\ae}bild, Domec, Tong, Brandt, et~al.]{tian2018coupling}
Feng Tian, Jean-Pierre Wigneron, Philippe Ciais, J{\'e}r{\^o}me Chave,
  J{\'e}r{\^o}me Og{\'e}e, Josep Pe{\~n}uelas, Anders R{\ae}bild,
  Jean-Christophe Domec, Xiaoye Tong, Martin Brandt, et~al.
\newblock Coupling of ecosystem-scale plant water storage and leaf phenology
  observed by satellite.
\newblock \emph{Nature ecology \& evolution}, 2\penalty0 (9):\penalty0
  1428--1435, 2018.

\bibitem[Todd et~al.(1998)Todd, Hoffer, and Milchunas]{todd1998biomass}
SW~Todd, RM~Hoffer, and DG~Milchunas.
\newblock Biomass estimation on grazed and ungrazed rangelands using spectral
  indices.
\newblock \emph{International journal of remote sensing}, 19\penalty0
  (3):\penalty0 427--438, 1998.

\bibitem[Tramontana et~al.(2016)Tramontana, Jung, Schwalm, Ichii, Camps-Valls,
  R{\'a}duly, Reichstein, Arain, Cescatti, Kiely,
  et~al.]{tramontana2016predicting}
Gianluca Tramontana, Martin Jung, Christopher~R Schwalm, Kazuhito Ichii, Gustau
  Camps-Valls, Botond R{\'a}duly, Markus Reichstein, M~Altaf Arain, Alessandro
  Cescatti, Gerard Kiely, et~al.
\newblock Predicting carbon dioxide and energy fluxes across global fluxnet
  sites with regression algorithms.
\newblock \emph{Biogeosciences}, 13\penalty0 (14):\penalty0 4291--4313, 2016.

\bibitem[Tsang et~al.(1985)Tsang, Kong, and Shin]{tsang1985theory}
Leung Tsang, Jin~Au Kong, and Robert~T Shin.
\newblock Theory of microwave remote sensing.
\newblock 1985.

\bibitem[Tsang et~al.(2000)Tsang, Chen, Chang, Guo, and Ding]{tsang2000dense}
Leung Tsang, Chi-Te Chen, Alfred~TC Chang, Jianjun Guo, and Kung-Hau Ding.
\newblock Dense media radiative transfer theory based on quasicrystalline
  approximation with applications to passive microwave remote sensing of snow.
\newblock \emph{Radio Science}, 35\penalty0 (3):\penalty0 731--749, 2000.

\bibitem[Ulaby et~al.(2014)Ulaby, Long, Blackwell, Elachi, Fung, Ruf,
  Sarabandi, Zebker, and Van~Zyl]{ulaby2014microwave}
Fawwaz~Tayssir Ulaby, David~G Long, William~J Blackwell, Charles Elachi,
  Adrian~K Fung, Chris Ruf, Kamal Sarabandi, Howard~A Zebker, and Jakob
  Van~Zyl.
\newblock \emph{Microwave radar and radiometric remote sensing}, volume~4.
\newblock University of Michigan Press Ann Arbor, 2014.

\bibitem[Vittucci et~al.(2019)Vittucci, Laurin, Tramontana, Ferrazzoli,
  Guerriero, and Papale]{vittucci2019vegetation}
Cristina Vittucci, Gaia~Vaglio Laurin, Gianluca Tramontana, Paolo Ferrazzoli,
  Leila Guerriero, and Dario Papale.
\newblock Vegetation optical depth at l-band and above ground biomass in the
  tropical range: Evaluating their relationships at continental and regional
  scales.
\newblock \emph{International Journal of Applied Earth Observation and
  Geoinformation}, 77:\penalty0 151--161, 2019.

\bibitem[Wang(1983)]{wang1983passive}
James~R Wang.
\newblock Passive microwave sensing of soil moisture content: The effects of
  soil bulk density and surface roughness.
\newblock \emph{Remote Sensing of Environment}, 13\penalty0 (4):\penalty0
  329--344, 1983.

\bibitem[Wei et~al.(2021)Wei, Qi, Ma, Wang, Ma, Gao, Huang, Zhao, Zhang, and
  Wang]{wei2021plant}
Da~Wei, Yahui Qi, Yaoming Ma, Xufeng Wang, Weiqiang Ma, Tanguang Gao, Lin
  Huang, Hui Zhao, Jianxin Zhang, and Xiaodan Wang.
\newblock Plant uptake of co2 outpaces losses from permafrost and plant
  respiration on the tibetan plateau.
\newblock \emph{Proceedings of the National Academy of Sciences}, 118\penalty0
  (33):\penalty0 e2015283118, 2021.

\bibitem[Wiesmann and M{\"a}tzler(1999)]{wiesmann1999microwave}
Andreas Wiesmann and Christian M{\"a}tzler.
\newblock Microwave emission model of layered snowpacks.
\newblock \emph{Remote sensing of environment}, 70\penalty0 (3):\penalty0
  307--316, 1999.

\bibitem[Wigneron et~al.(2008)Wigneron, Chanzy, De~Rosnay, Rudiger, and
  Calvet]{wigneron2008estimating}
Jean-Pierre Wigneron, Andre Chanzy, Patricia De~Rosnay, Christoph Rudiger, and
  Jean-Christophe Calvet.
\newblock Estimating the effective soil temperature at l-band as a function of
  soil properties.
\newblock \emph{IEEE Transactions on Geoscience and Remote Sensing},
  46\penalty0 (3):\penalty0 797--807, 2008.

\bibitem[Wigneron et~al.(2021)Wigneron, Li, Liu, Wang, Frappart, Fan, Al-Yaari,
  Fernandez-Moran, Ma, Ygorra, et~al.]{wigneron2021alternate}
Jean-Pierre Wigneron, Xiaojun Li, Xiangzhuo Liu, Mengjia Wang, Fr{\'e}d{\'e}ric
  Frappart, Lei Fan, Amen Al-Yaari, Roberto Fernandez-Moran, Hongliang Ma,
  Bertrand Ygorra, et~al.
\newblock Alternate inrae-bordeaux vod indices from smos, amsr2 and ascat:
  Overview of recent developments.
\newblock In \emph{2021 IEEE International Geoscience and Remote Sensing
  Symposium IGARSS}, pages 6210--6213. IEEE, 2021.

\bibitem[Wild et~al.(2022)Wild, Teubner, Moesinger, Zotta, Forkel, van~der
  Schalie, Sitch, and Dorigo]{wild2022vodca2gpp}
Benjamin Wild, Irene Teubner, Leander Moesinger, Ruxandra-Maria Zotta, Matthias
  Forkel, Robin van~der Schalie, Stephen Sitch, and Wouter Dorigo.
\newblock Vodca2gpp--a new, global, long-term (1988--2020) gross primary
  production dataset from microwave remote sensing.
\newblock \emph{Earth System Science Data}, 14\penalty0 (3):\penalty0
  1063--1085, 2022.

\bibitem[Wohlfahrt et~al.(2008)Wohlfahrt, Hammerle, Haslwanter, Bahn,
  Tappeiner, and Cernusca]{wohlfahrt2008seasonal}
Georg Wohlfahrt, Albin Hammerle, Alois Haslwanter, Michael Bahn, Ulrike
  Tappeiner, and Alexander Cernusca.
\newblock Seasonal and inter-annual variability of the net ecosystem co2
  exchange of a temperate mountain grassland: Effects of weather and
  management.
\newblock \emph{Journal of Geophysical Research: Atmospheres}, 113\penalty0
  (D8), 2008.

\bibitem[Xu et~al.(2021)Xu, Saatchi, Yang, Yu, Pongratz, Bloom, Bowman, Worden,
  Liu, Yin, et~al.]{xu2021dataset}
Liang Xu, Sassan~S Saatchi, Yan Yang, Yifan Yu, Julia Pongratz, A~Anthony
  Bloom, Kevin Bowman, John Worden, Junjie Liu, Yi~Yin, et~al.
\newblock Dataset for" changes in global terrestrial live biomass over the 21st
  century".
\newblock 2021.

\end{thebibliography}

\end{document}